\documentclass{aa}

\usepackage{txfonts}
\usepackage{graphicx}
\usepackage{amssymb,amsmath}
\usepackage{layout}
\usepackage{graphics}
\usepackage{hyperref}

\begin{document}

   \title{Morphologies arising from the gas flow in the innermost kiloparsec of barred galaxy models}

   \subtitle{ }

   \author{S.~Pastras\inst{1,2}
      \and P.A.~Patsis\inst{3,2}
      \and E.~Athanassoula\inst{4}
          }

   \institute{Max-Planck-Institut f\"{u}r Extraterrestrische Physik (MPE), Gießenbachstr. 1, D-85748 Garching, Germany\\
              \email{spastras@mpe.mpg.de}
         \and Max-Planck-Institut f\"{u}r Astrophysik (MPA), Karl-Schwarzschild-Str. 1, D-85748 Garching, Germany
         \and Research Center for Astronomy and Applied Mathematics, Academy of Athens, Soranou Efessiou 4, 11527 Athens, Greece\\
              \email{patsis@academyofathens.gr}
         \and Aix Marseille Universit\'{e}, CNRS, LAM (Laboratoire d' Astrophysique de Marseille), UMR 7326, 13388 Marseille 13, France\\
              \email{evangelie.athanassoula@lam.fr}
             } 

   \date{Received xxxx; accepted xxxx}

   \abstract
   {We study a series of response models to investigate the formation of specific morphological features in the central 1 kpc region of the gas component in barred spiral galaxies.}
   {We aim to understand how structures, such as nuclear rings and spirals, form by
varying the parameters of a general gravitational potential and gas
properties. Our goal is to determine how much the shape of these structures is
driven by the orbital dynamics of the models compared to the influence of the
hydrodynamics of the gas. In particular, we examine the effects of the bar
strength, bar shape, pattern speed, and central density, as well as their mutual
interdependence.}
   {We modeled the gas flow using hydrodynamical simulations run with the Eulerian
\texttt{RAMSES} code. The underlying gravitational potential was a two-dimensional Ferrers
bar and the gas was considered to be isothermal. Alongside analyzing the gas response
to the imposed gravitational potentials, we carried out orbital studies for all
models. This involved assessing the shapes and stability of periodic orbits and
analyzing the distribution of regular versus chaotic regions within the
systems.}
   {The parameters of the gravitational potential alone are insufficient to
accurately predict the gas dynamics in a system. The morphology of the gaseous
response varies substantially with changes in sound speed, emphasizing the
fundamental role of hydrodynamic processes in determining the structure of the
gas within the central region. We identify the factors that affect the
morphology of nuclear rings and trailing and leading nuclear spirals.
The best alignment between our models and structures observed in local barred
galaxies is achieved by assuming a sound speed of $c_s=20\,\rm{km\,s^{-1}}$.}
   {}


   \maketitle

\nolinenumbers


\section{Introduction}
\label{sec:introduction}

The innermost kiloparsec of a galactic bar is its central region, which in 
some cases includes an area encircled by a nuclear ring \citep[see for 
example][]{cometal10} along with the nuclear ring itself. Observations provide
sufficient information about the topology and the general 
morphology of nuclear rings, allowing even for their classification into 
subclasses \citep{bc96, cometal10}.

However, the area within these rings is less explored. The primary challenges for observing structures
within the innermost kiloparsec are resolution and projection effects. There are 
significantly fewer studies that attempt to identify common morphological 
features in the central regions of galaxies and connect them to the dynamics of 
the parent bar.

Notably \citet{mrmp03, mrmp03b} used \textit{HST} images of an extensive galaxy sample to 
map the distribution of the cold interstellar medium in the central parts of disk 
galaxies using dust as a tracer. Among the observed morphologies, they 
identified nuclear dust spirals, which could be categorized as grand design, 
tightly wound, loosely wound, or chaotic (flocculent). In certain instances, 
they also detected chaotic circumnuclear dust within the innermost kiloparsec, or, in 
some cases, the absence of any circumnuclear dust structure. Their findings 
indicated a connection between grand design nuclear spirals and large-scale 
bars. Contrarily, they found that tightly wound nuclear spiral arms are more
commonly observed in unbarred galaxies rather than in barred ones.

Furthermore, \citet{mrmp03, mrmp03b} revealed that the majority of barred 
galaxies do not exhibit rings inside the bar, implying that nuclear spirals are 
often not situated within a nuclear ring. In such instances, they are linked directly 
to the relatively straight dust lanes that are associated with the large-scale 
bar. In such cases, the dust lanes essentially form continuous structures spanning
from scales of several kiloparsecs down to within tens of parsecs from the galactic 
center. Such an example is that of NGC~1530 \citep[see for example][- 
figure 1-25]{butarev}, in which a curve is observed in the primary dust lanes of the bar near the center, without the clear presence of a nuclear ring.

Nevertheless, the regions inside the rings are not always featureless. An 
example of grand design spiral arms inside a nuclear ring is observed in the 
recently released \textit{JWST} image of NGC~1433. \citet{pxa21} identified a nuclear 
trailing spiral embedded within the peanut-shaped bulge of NGC~352. Another 
example, this time of a multiarmed spiral, is identified inside the nuclear 
ring of NGC~1512 as observed by the Hubble space telescope. There are also cases in which 
the nuclear spiral has three arms, as in the case of NGC~1097, where a 
three-armed spiral can be identified inside a nuclear ring \citep{pmr05, kf06, Davies_2009}
(however, see also \citealt{Kolcu_2023}).

Regarding the nuclear rings themselves, in numerous cases they exhibit a 
pseudo-ring nature, appearing as an extension of the primary dust lane shocks. 
The surface density is not uniform along a ring and the denser segments are
located in the quadrants that extend inward from the primary shocks.
Because of this, they can be referred to as pseudo-rings. Typical 
examples can be observed, for instance, in the SDSS image of NGC~5383, in the ring 
of NGC~4736 \citep{vdL15}, or in the \textit{HST} image of NGC~1512. This class of 
pseudo-rings can be viewed as displaying a morphology that lies between those 
with well-defined nuclear rings and those where the main dust lane shocks curve 
inward, extending into the region within 1 kpc.

Another morphological feature associated with the central regions of barred 
galaxies, and predicted by the gaseous density wave theory in its linear 
approximation \citep{gt78, gt79}, is the formation of leading spirals. They are 
expected to appear at the inner inner Lindblad resonance (iILR) region 
propagating  outward \citep[see for example][]{wm04a}, provided that this 
resonance exists. Nevertheless, there are not many known examples of leading 
spirals in the innermost kiloparsec regions of galactic disks. We mention the 
K$^{\prime}$ image of NGC~6902  \citep{pg03}, which however is a Sa(r) type 
galaxy. \citet{dz03} reported the presence of a well-defined two-armed leading 
spiral inside a nuclear ring observed in Pa$\alpha$ emission at the core of 
NGC~1241. In this case, farther inward, a tiny bar could also be identified.

Central leading spirals have been simulated in analytic models and models of the 
gas response to bar potentials, under various assumptions. These spirals are 
anticipated to manifest in the gaseous orbits of weak bar potentials, irrespective
of whether the effects of self-gravity are considered \citep{w94, wk01, wm04a}.
They have been reproduced in hydrodynamical simulations by \citet{at05} using
smoothed particle hydrodynamics (SPH), by \citet{kss12, Kim_2012} using the grid-based code
\texttt{CMHOG} \citep{pin95}, by \citet{sbm15II} by means of a second-order flux-splitting scheme
\citep{vA82,ath92b} and by \citet{lsk15} using the grid-based magnetohydrodynamics code
\texttt{Athena} \citep{Stone_2008}, which utilizes a higher-order Godunov scheme \citep{sg09}.

The presence of leading nuclear spirals in models with ILRs is claimed
to be linked to the variation in the precession rate of elliptical periodic
orbits. They emerge when the shape of the $\Omega - \kappa/2$
curve in its innermost part is decreasing toward the center. Here, $\Omega$
represents the angular frequency and $\kappa$ denotes the epicyclic frequency.
On the contrary, if this curve increases toward the center, for instance, due to the presence
of a central massive black hole (MBH), the gas forms a trailing spiral. This
scenario is aptly elucidated in \citet[][- figure 79]{bc96} and \citet[][-
figure 2]{fc22}.
 
It is puzzling that nuclear leading spirals are rarely observed in images 
of barred spiral galaxies. This is surprising considering that in the typical 
cases of standard barred galaxy potentials, the $\Omega - \kappa/2$ curves tend 
to decrease toward the center, allowing the coexistence of an iILR and an outer 
ILR (oILR) \citep[see e.g.,][- figure 1]{gco80} for a wide range of realistic 
bar pattern speeds. Nevertheless, it is indeed true that, despite being 
predicted in some cases, nuclear leading spirals do not form in all hydrodynamic
simulations. This discrepancy may stem either from the extent of the trailing
spiral produced at the oILR, numerical issues related to the simulation itself,
such as the resolution of the numerical scheme employed \citep{wm04b}, or
because they are simply a transient feature \citep{at05}.

In this study, we investigate the circumstances that lead to the
appearance of all these morphological features within the central regions of the
gaseous components of barred galaxies in models that do not explicitly include a
black hole, represented by a point mass. Several numerical studies of bar-induced
flows in isothermal gas disks have examined how the resulting structures depend
on the adopted sound speed, using a variety of different imposed potentials
\citep[e.g.,][]{eg97, pa00, wm02, skr03, kss12, sbm15I}. Other works focus on the effects
of various potential parameters, but do not additionally explore the dependence
of the resulting morphologies on the sound speed of the gas \citep[e.g,][]{rt03, Kim_2012}.
Unlike previous works, our study systematically investigates the effects of multiple
parameters on the formation of nuclear spirals, all within a single, general
Ferrers bar potential, and in each case for an array of different gas sound speeds.
By varying only one parameter at a time while keeping the others fixed, we are able
to isolate its individual influence on the gas response in the innermost kiloparsec.

Our objective is to understand how variations in the underlying gravitational
potential and gas properties affect the formation of nuclear rings and nuclear spirals.
We aim to determine the extent to which the resulting shapes of such structures are
influenced by the orbital characteristics of our models versus the hydrodynamics of the gas.
Particular emphasis is put on less explored factors, such as the influence of the bar shape,
bar strength, pattern speed, and central density of the imposed potential. We also focus on
investigating how the sound speed governs the emergence of leading spirals within nuclear rings.

We use a grid-based hydrodynamical code with a spatial resolution high
enough to resolve structures within the central kiloparsec, allowing us to
capture the nuclear features in detail. This represents a complement to the
classic study by \citet{ath92b}, which successfully resolved shocks in the main
bar region but did not examine the gas response in the central parts of the
models.

In Section~\ref{sec:models}, we provide details regarding the general potential 
used, the specific cases we analyzed, the hydrocode employed, and the way we 
approach the stellar dynamics of the examined cases. Subsequently, 
in Section~\ref{sec:results}, we present the orbital content of the models and 
the response of the gas in the five cases we have considered. To assess the 
influence of the bar component in our models, we compare the primary results 
from Section~\ref{sec:results} with models of extremely weak bars in Section~\ref{sec:axisym}.
Section~\ref{sec:dis} contains our discussion and summarizes our conclusions.

\section{The models we studied}
\label{sec:models}

\subsection{The gravitational potential}
\label{sec:gravitationalPotential}

In our study, we adopted the general gravitational potential used by
\citet{ath92a}, which encompasses three components representing a bulge, a disk, 
and a bar. The bulge has a volume density of
\begin{equation}
\rho(r)=\rho_b \left( 1+\dfrac{r^2}{r{_b}^2}\right) ^{-3/2},
\end{equation}
with $\rho_b$ and $r{_b}$ being, respectively, the bulge central density and scale length.

The surface density of the disk is given by 
\begin{equation}
 \sigma(r)=\dfrac{v_0^2}{2\pi Gr_0} \left(1+\dfrac{r^2}{r{_0}^2}\right) ^{-3/2},
\end{equation} 
with $v_0$ and $r{_0}$ being scaling constants, i.e., it is a Kuzmin-Toomre disk 
\citep{ku56, too63}.

The bar is a standard Ferrers ellipsoid \citep{f1877} with volume 
density
\begin{equation}
 \rho_B(x,y,z)= \left\{
 \begin{array}{ll}
  \rho_0 (1-g^2) & \text{for $g<1$}\\
  0 & \text{elsewhere,}\\
 \end{array}
\right. 
\label{eq:rhoBar}
\end{equation} 
with \begin{displaymath}
      g^2 = \frac{y^2}{a^2}+\frac{x^2+z^2}{b^2}, \text{and}\; a>b, \text{with}\; a=5\,{\rm kpc}.
     \end{displaymath}

The quadrupole moment, $Q_m$, of the bar along its major axis, $q_y$, is given by 
\begin{equation}
\label{eq:qm}
 q_y = 2 M_B (a^2 - b^2)/7 = 2 Q_m,
\end{equation} 
where $M_B$ is the mass of the bar.

The units used for the model parameters are, $10^6 {\rm M}_{\odot}$ for the 
mass, 1\,kpc for the length, and $1\,{\rm km\,s^{-1}}$ for the velocity. The resulting 
angular frequency, $\Omega$, unit is $1\,{\rm km\, s^{-1}\, kpc^{-1}}$. 
For further details the reader may refer to \citet{ath92a}.

This model enables us to explore the dynamics of bars in their central regions 
for different parameter sets. Its straightforward nature facilitates the
correlation of particular dynamic behaviors with variations in specific 
parameters, thereby assessing the significance of these parameters in relation 
to the emergence of specific morphological features. Moreover, this potential has been
thoroughly studied in the literature, making direct comparisons with previous research possible.
 
Specifically, we examined five models characterized by typical bar parameters. 
Four of these are listed in \citet[][- table 1]{ath92a}. We chose to keep the same
reference numbers for these four cases. The fifth model (R901) was included to explore
a different case with a noticeably unique shape in its rotation curve compared to the others.

The parameters characterizing these models are summarized in Table~\ref{tab:mparams}.
In the table, we give, from left to right, the name of the model, the bar 
axial ratio $a/b$, the radius of the $L_1$ and $L_2$ Lagrangian points, a measure
of the quadrupole moment of the bar along its major axis, $Q_m$ (see Eq.~\ref{eq:qm}),
the sum of the bar and bulge central density, $\rho_c = \rho_0 + \rho_b$, and the
pattern speed of the bar, $\Omega_B$.

\begin{table}[h]
\caption{Dynamical parameters of the five main models.}
\label{tab:mparams}
\centering
\begin{tabular}{c c c c r c}
\hline\hline\\ [-1.5ex]
Model & $a/b$ &  $r_L$ & $Q_m/10^4$ & $\rho_c/10^4$ & $\Omega_B$ \\
\hline\\ [-1.5ex]
   R001 & 2.5 & 6.0 & 4.5 & 2.4  & 35.300\\
   R010 & 1.5 & 6.0 & 4.5 & 2.4  & 35.437\\
   R153 & 2.5 & 3.0 & 4.5 & 2.4  & 77.726\\
   R192 & 2.5 & 6.0 & 1.5 & 2.4  & 33.662\\
   R901 & 2.5 & 6.0 & 4.5 & 20.0 & 35.615\\
\hline
\end{tabular}
\end{table}

Moreover, we examined two additional cases, with parameters resembling those of 
models R001 and R901, except that these models featured an extremely weak
bar, effectively making them nearly axisymmetric (as we explain in
Section~\ref{sec:axisym}). This was carried out to assess the significance of the
gravitational forces relative to the hydrodynamical characteristics of the gas 
in determining the fundamental conclusions of our study.

\subsection{Modeling the gas flow}
\label{sec:ramses}

To model the gas flow, we conducted a series of hydrodynamic simulations using 
the Eulerian \texttt{RAMSES} code \citep{ramses1}, which has been effectively applied 
in similar gas dynamical studies of both barred and non-barred spiral galaxies 
\citep[see e.g.,][]{few16, ff16}. Gas dynamics in \texttt{RAMSES} are simulated using a
second-order unsplit Godunov scheme and an adaptive mesh refinement technique. 
In our models, the maximum achieved resolution is $\approx10$ pc, 
much higher than in \citet{ath92b}. In all cases, the gas is considered 
isothermal with an adiabatic index of 5/3. For each one of our five bar models 
(R001, R010, R153, R192, and R901) we run three simulations with sound speeds 
$c_s=2\,\rm{km\,s^{-1}}$, $10\,\rm{km\,s^{-1}}$, and $20\,\rm{km\,s^{-1}}$, 
respectively.

The code was tested by \citet{sp22a} for the types of simulations presented in this work, specifically,
for all Ferrers models with an index of n=1 in \citet{ath92a}. All
responses yielded identical results to the corresponding models in
\citet{ath92b} in terms of morphology and provided a very similar description of
the dust lane shocks. The \texttt{RAMSES} responses were also compared with SPH \citep{gm77, ll77} models
for the Ferrers bars and for a model of NGC~7479 \citep{ppa22}. The main morphological
features and the corresponding flows were consistent in the responses of the two codes.
Differences observed in the details of the developed morphological features, as well as the advantages
and disadvantages of each code, are thoroughly discussed in \citet{sp22a} and
\citet{ppa22}.

In our models, the gas responds to the potential described in Section~\ref{sec:models} for a 
time, $t$, corresponding to approximately 15 pattern rotations. The initial gas 
surface density is the same as in \citet{ath92b}, i.e., 1~M$_{\odot}$~pc$^{-2}$.
In the absence of self-gravity, as is the case for our models, this value is effectively a scaling factor.

The simulations begin with an initial growing bar phase that lasts for a time,
$t_f$,  corresponding to the time for three pattern rotations. During $t_f$, the 
mass of an axisymmetric disk with mass $M_B$ is gradually transferred in a 
linear manner to the Ferrers bar component. In other words, during $t_f$, a 
potential component with a density of the form of Eq.~\ref{eq:rhoBar},
but with $a\approx b$, is transformed into a Ferrers bar in which $a > b$.

Finally, it is worth noting that while all our responses are in principle
symmetric with respect to the origin, there are some slight deviations
from perfect symmetry, possibly due to local hydrodynamical instabilities.
While they do not affect any of the main conclusions of this work, traces
of these asymmetries can be identified in the low density regions of the
presented slit profiles.

\subsection{Stellar dynamics}
\label{sec:orbits}

One of our objectives is the identification of all factors influencing
the eventual shapes of the structures in the innermost kiloparsec. Thus,
for each model, in addition to examining how the gas responds to the imposed
potential, we conducted an orbital analysis following the procedures outlined in
\citet{ath92a}. These calculations were performed in the frame of reference that
rotates with $\Omega_B$. The equations of motion are derived from the Hamiltonian,

\begin{equation}
 H = \frac{1}{2}\left(\dot{x}^2 +\dot{y}^2\right) + \Phi(x,y) - 
\frac{1}{2} \Omega_B^2 \left(x^2 + y^2\right),
\end{equation}

where $(x, y)$ are the coordinates in a Cartesian frame of reference corotating 
with the bar, with angular velocity $\Omega_B$. $\Phi(x,y)$ is the potential 
(see Sect.~\ref{sec:models}) in Cartesian coordinates, $E_J$ is the numerical 
value of the Jacobian integral, and dots denote time derivatives. The unit of $E_J$
is $1\,{\rm km^2\, s^{-2}}$. For the integration of the orbits we used a 
fourth-order Runge-Kutta integration scheme with a variable step. 
Furthermore, when necessary, we assessed the stability of the periodic orbits
by analyzing the variation in H{\'e}non's index, as described in \citet{henon}. 

In our work, the location of the ILR resonance region is crucial, since
it defines the domain of what we call the ``x2-flow''\footnote{Note
that orbits in the ILR region must have the appropriate Jacobi constants in
order to belong to the x2-flow.}. We associate the positions of the iILR
and oILR with the range spanned by the innermost and outermost periodic orbits
within the x2-x3 loop in the characteristic diagrams or in H{\'e}non's
stability diagrams (see Section~\ref{sec:r001orbs}), rather than with radii
derived from frequency curves for a given pattern speed, based on the
axisymmetric component of the potential, or with any other alternative
definition of an ILR radius (but see also \citealt{rt03}).

Stellar response models were also implemented by adopting a procedure akin 
to that used for the gaseous models. As for the latter, starting from a purely
axisymmetric background, we incremented the perturbation until the desired strength was reached. 
Subsequently, we continued integrating the initial conditions for a duration 
corresponding to an additional ten pattern rotations. These stellar models are useful 
as they offer insights into the orbital families that ultimately become populated.

\section{Results}
\label{sec:results}

\subsection{Model R001 - The fiducial case}
\begin{figure}[h]
    \centering
	\includegraphics[width=\columnwidth]{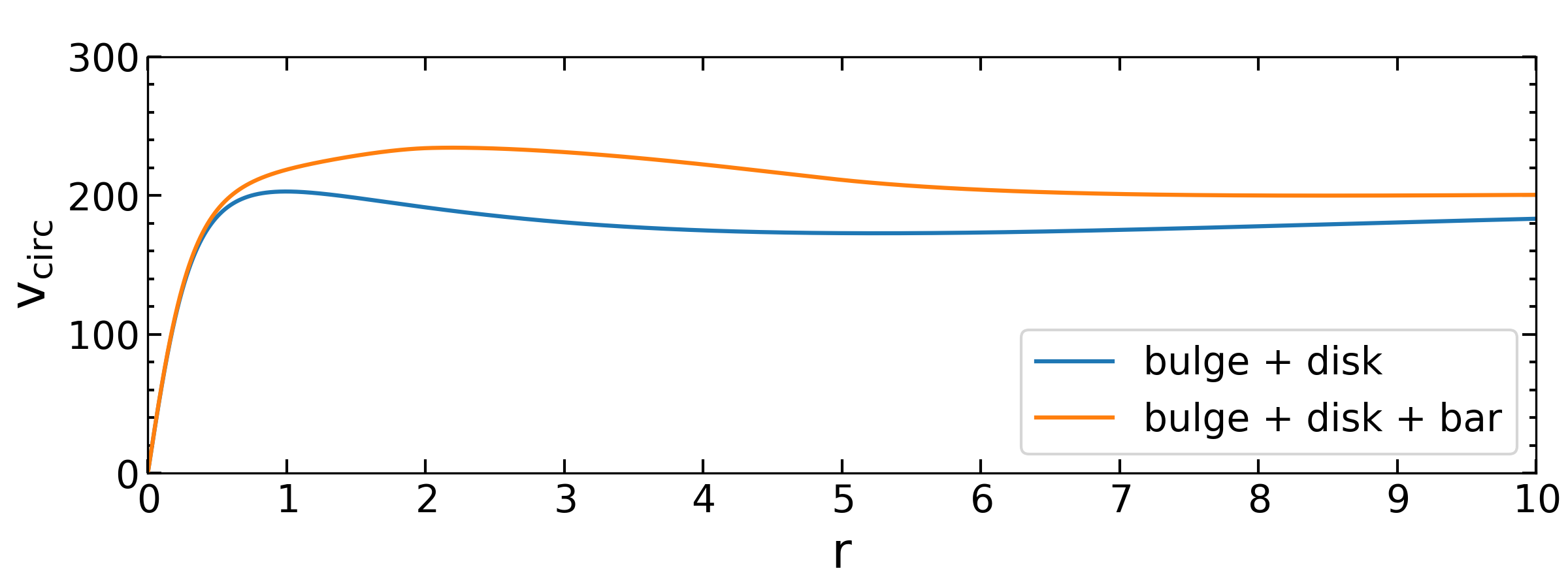}
    \caption{Rotation curve of model R001. The blue curve corresponds to the axisymmetric part of the potential, while the orange one to the full potential.}
    \label{fig:rc_krun1}
\end{figure}

We use model R001, which corresponds to RUN 1 in \citet{ath92a}, as our fiducial model.
The rotation curve of the model is given in Fig.~\ref{fig:rc_krun1}.

The curve representing the axisymmetric component of the potential is shown in 
blue, while the angle-averaged total rotation curve is illustrated in orange. 
This rotation curve is characteristic of a typical disk galaxy.

\subsubsection{Model R001 - Orbital dynamics}
\label{sec:r001orbs}

The primary orbital characteristics of this model have been presented in detail
in \citet{ath92a}. In this context, we highlight specific aspects that are crucial 
for comprehending the dynamics of the gas within the innermost 1~kpc region.
 
\begin{figure}
    \centering
	\includegraphics[width=\columnwidth]{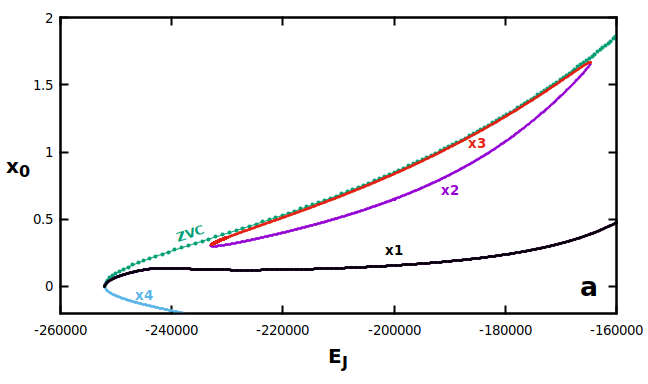}
    \includegraphics[width=\columnwidth]{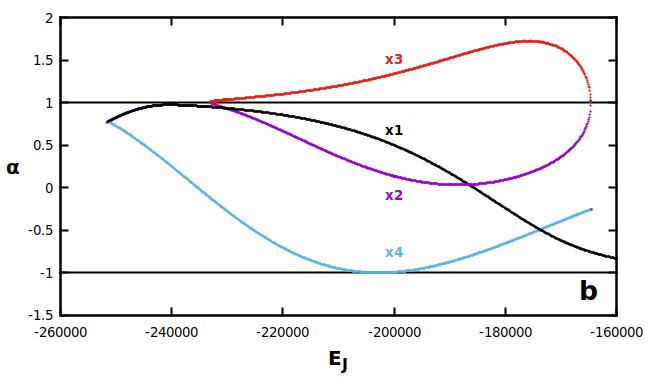}
    \caption{(a) Characteristic diagrams for the x1, x2 and x4 families in model 
R001. The zero velocity curve is depicted with a green color and indicated with 
``ZVC.'' (b) H\'{e}non index, $\alpha$, for the same families. We find no changes
in the stability of any of these families.}
    \label{fig:char_krun1}
\end{figure}

Within the central region of this model, we only encounter x1, x2-x3, and x4 
orbits\footnote{We note that this is not the general case. For instance, in 
some potentials derived from near-infrared observations, we might observe 
a complex evolution of the standard families with bifurcations in the innermost 
kiloparsec region, along with orbits linked to the 1:1 resonance \citep[refer to, for 
example,][]{paq97}.}. The characteristic diagrams \citep[see Sect.~2.4.4 
of][]{gcobook} of these families for Jacobi constants, $E_J$, ranging from the 
center of the system up to the oILR of the model, are depicted in
Fig.~\ref{fig:char_krun1}a. The existing families are indicated with labels
with the same color as the corresponding characteristic curves, while the
zero velocity curve (ZVC) for $x > 0$ is also given with a dotted green line. 
 
The H\'{e}non indices of the same families in the same $E_J$ range are given in 
Fig.~\ref{fig:char_krun1}b. The stability of the families remains 
unchanged within this energy interval. Although there is a local maximum on the 
x1 stability curve at $(-240800, 0.97)$, there is no intersection or tangency 
with the $\alpha = 1$ axis. Consequently, the x1 and x2 families do not give 
rise to bifurcating families within the system. Also, the tangency of the x4 
stability curve with the $\alpha = -1$ axis at approximately $-202900$ is not 
expected to significantly influence the local or global dynamics of 
the system, as it corresponds to retrograde periodic orbits, not populated in the 
model.

\begin{figure}
    \centering
	\includegraphics[width=\columnwidth]{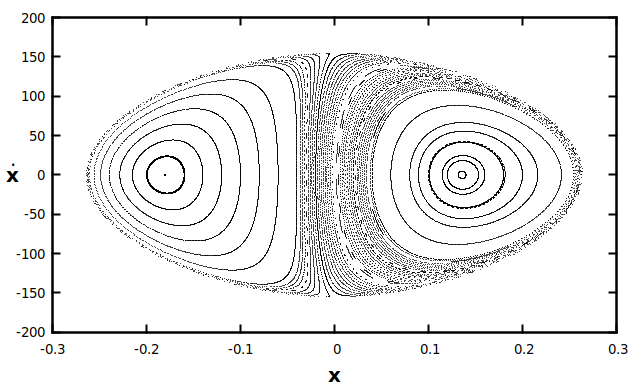}
    \caption{Typical surface of section for $E_J < E_J(\text{iILR})$ of model R001,
at $E_J=-240000$. The x4 periodic orbit is located at the center of the
left stability island, while x1 is at the center of the right one.}
    \label{fig:sos24}
\end{figure}

Essentially there is no chaos at the innermost kiloparsec region of the model. A 
typical $(x,\dot{x})$ surface of section (sos) for $E_J < E_J(\text{iILR})$ 
is shown in Fig.~\ref{fig:sos24}. In this representation, practically we can 
exclusively identify invariant curves. Order also dominates in the surfaces of 
section for $E_J(\text{iILR})<E_J<E_J(\text{oILR})$ (not given in a separate 
figure here), featuring a third stability island that emerges around $x_0$(x2) 
$> x_0$(x1) and negligible chaotic zones, mainly due to the presence of x3, 
which is situated between the boundaries of the x2 stability island and the 
ZVC. 

\begin{figure}
    \centering
	\includegraphics[width=0.8\columnwidth]{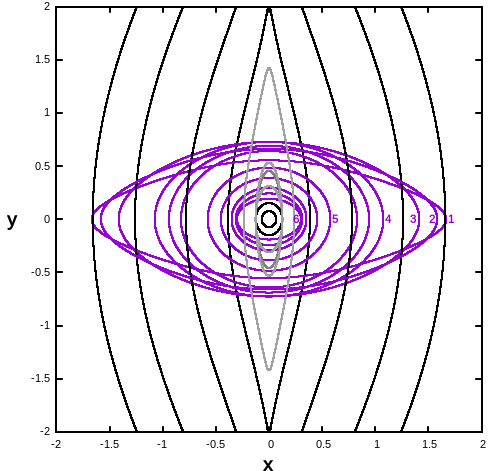}
    \caption{Orbits of the x1 and x2 families in model R001. The flow in our 
response model follows either the black x1, or the magenta x2 orbits. The gray 
x1 orbits exist for $E_J$'s, where only x2 orbits are populated. The labels from 
``1'' to ``6'' indicate successively the x2 orbits in descending $E_J$ order.}
    \label{fig:krun1_orbs}
\end{figure}

The orbital background of the stellar response model is composed by the 
standard elliptical-like pro-grade x1 and x2 periodic orbits. In the manner in 
which we construct our models, beginning with circular motion in the 
axisymmetric background and gradually increasing the mass of the bar component 
(refer to Section~\ref{sec:orbits}), at each $E_J$, the flow is influenced by 
either x1 or x2 orbits. We find that the orbits populating the model 
are linked to the family with initial conditions closer to the circular orbits 
of the axisymmetric background at each $E_J$. The periodic orbits of these 
families are depicted in Fig.~\ref{fig:krun1_orbs}. In the plot, the x1 orbits are
given in black and the x2 orbits in magenta. The orbits in gray correspond to 
x1 orbits that we found not to be populated, at $E_J$ values where the flow 
adheres to the x2 pattern (see Section~\ref{sec:gasres} below).

The outermost x1 periodic orbit in Fig.~\ref{fig:krun1_orbs}, occurring at $E_J = 
-134097.8$, being tangent to the outermost x2 orbit at $E_J = -164591.4$, is 
essentially the outermost elliptical-like x1 orbit of the model. This is 
because, for larger $E_J$ values, the x1 orbits undergo a relatively abrupt 
transition to a rectangular-like shape as they approach the 4:1 resonance 
region. The abruptness of the transition can be understood in the increment of  
the $x_0$ initial condition of x1 from 1.66 at $E_J = -134097.8$ to 2.62 at
$E_J = -130000$, pushing it well away from the x2 region 
\citep[cf. with figure 2 in][]{ath92a}. 
 
As is seen in Fig.~\ref{fig:krun1_orbs} there is a crowding of x2 orbits in the 
direction of the major axis of the bar (y axis) at about 0.7~kpc, as expected   
\citep[][- figure 6]{gco79}. Actually there are even intersections of x2 orbits 
among themselves in a narrow $E_J$ interval. In Fig.~\ref{fig:krun1_orbs} the labels 
of the x2 orbits indicate successive periodic orbits in a descending $E_J$ order.
We find that orbit ``5,'' for  $E_J = -205000$, is the outermost x2
one that is not intersecting other members of the same family. All other x2 periodic orbits,
for larger $E_J$, intersect other members of the x2 family.

Crowding of orbits is observed not only at the outer boundary of the x2 region, 
but at its inner boundary as well, as the ellipticity of the x2 orbits increases 
again, eventually reaching the shape of the x2 periodic orbit labeled ``6.'' The crowding 
of orbits at this region is also conspicuous. For even smaller $E_J$ than that 
of orbit ``6,'' we encounter again orbits of the x1 family, of low ellipticity, 
which at the very central region are almost circular. Specifically, as $E_J$ 
decreases to around $-223000$, the x1 orbits begin to resemble ellipses  with 
low ellipticity, generally extending to sizes of $r \leq 150$\,pc 
(Fig.~\ref{fig:krun1_orbs}).
 
The response models, during the initial transient phase, lasting for the first
three pattern rotations, essentially determine which orbital families will be
populated. Given that the potential variation is the
same in both stellar and gaseous models, understanding the orbital
families present in the stellar model helps in comprehending the flow
within the corresponding regions in the gas case.
This will become apparent in the subsequent sections.

\subsubsection{Model R001 - Gaseous response}
\label{sec:gasres}

\begin{figure*}
    \centering
	\includegraphics[width=2\columnwidth]{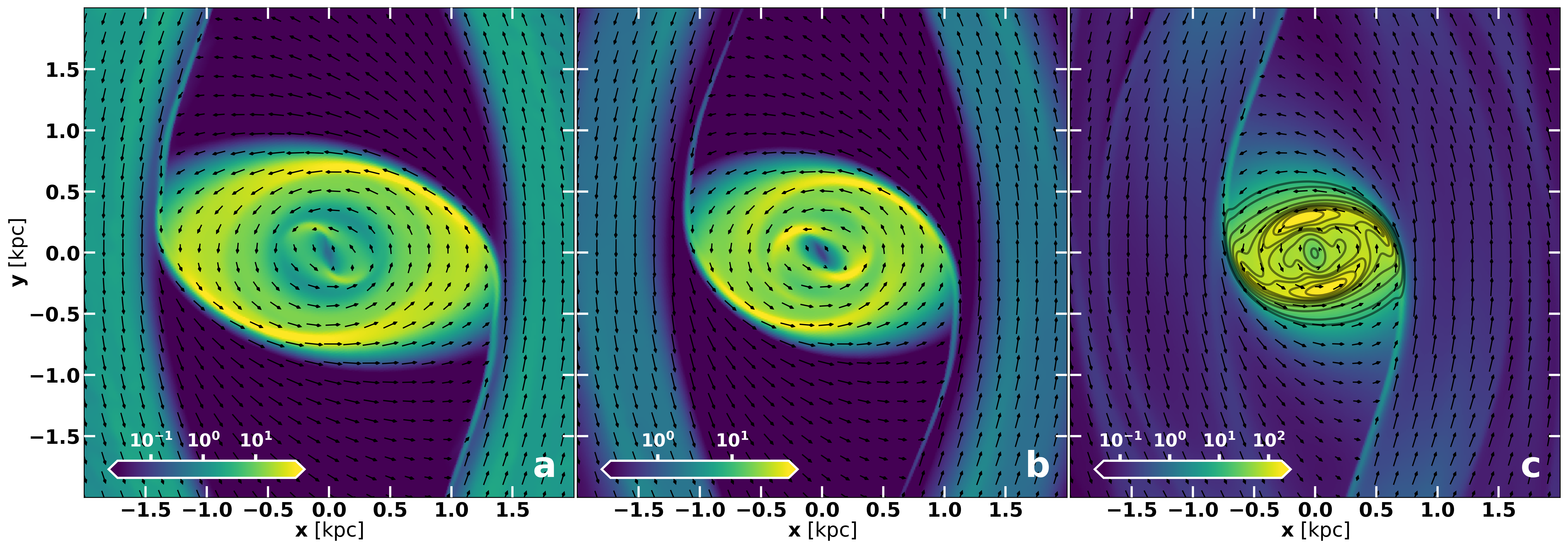}
    \caption{Model R001: Central regions of three models with different sound
speeds, (a) $c_s=2$~km\,s$^{-1}$, (b) $c_s=10$~km\,s$^{-1}$,
and (c) $c_s=20$~km\,s$^{-1}$. A leading spiral is apparent in (a) and 
(b), while in (c) we have the formation of a smaller pseudo-ring.}
    \label{fig:r1l2}
\end{figure*}

The morphology of the response in the central 4$\times$4~kpc region of model 
R001 at time $t=2.99$, equivalent to approximately 16.8 dynamical times (or 13.8 
with the full potential) is depicted in Fig.~\ref{fig:r1l2}. Successively we give the 
responses for the models with $c_s=2\,\rm{km\,s^{-1}}$ (Fig.~\ref{fig:r1l2}a),  
$c_s=10\,\rm{km\,s^{-1}}$ (Fig.~\ref{fig:r1l2}b), and $c_s=20\,\rm{km\,s^{-1}}$ 
(Fig.~\ref{fig:r1l2}c). Clearly, we observe the development of a noticeable leading 
spiral feature for sound speeds of $c_s=2\,\rm{km\,s^{-1}}$ and $c_s=10\,\rm{km\,s^{-1}}$
(Fig.~\ref{fig:r1l2}a and Fig.~\ref{fig:r1l2}b, respectively).

In the $c_s=2\,\rm{km\,s^{-1}}$ case, the orientation of the leading spiral arms remains 
consistent throughout the evolution for $t > t_f$. Two visibly open spiral arms 
develop from the extremities of the major axis of a central, oval-shaped, 
region, tilted at approximately 72\textdegree~relative to the minor axis of the 
bar (x axis), toward its leading side. The color coding on the 
bottom-left side of the figures suggests that this central, oval area is 
a region of low surface density, significantly less dense than its immediate 
surroundings. The dense leading spiral arms emerge near the extremities of the 
major axis of this oval structure. The overall morphology is similar to that
of the model without a black hole by \citet[][- figure 6]{at05}.

\begin{figure*}
    \centering
	\includegraphics[width=2\columnwidth]{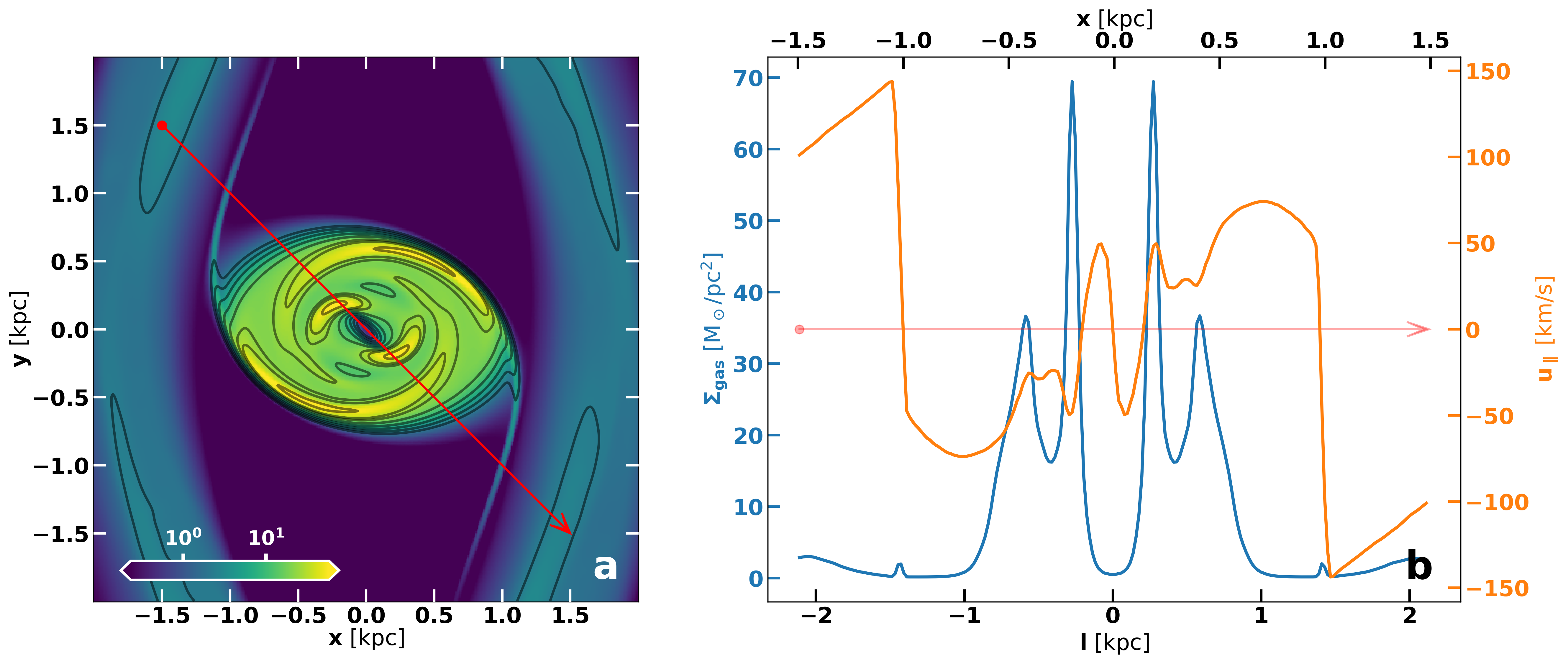}
    \caption{Model R001: Variation in gas surface density ($\Sigma$) and the 
velocity component along a defined slit ($u_{\parallel}$) for $c_s$=10~km\,s$^{-1}$. 
Panel (a) shows the slit’s location, while panel (b) presents the variations, 
with the blue curve indicating $\Sigma$ and the orange curve representing 
$u_{\parallel}$.}
    \label{fig:lineR1_10_300a}
\end{figure*}

In Fig.~\ref{fig:r1l2}b, for $c_s=10\,\rm{km\,s^{-1}}$, the orientation of
the oval has shifted, forming an angle of approximately 50\textdegree~relative
to the minor axis of the bar, once again toward the leading side
of the bar. Remarkably, the pseudo-ring structure previously seen at
the outer border of the x2 region in Fig.~\ref{fig:r1l2}a with
$c_s=2\,\rm{km\,s^{-1}}$, has transformed into a trailing spiral
pattern that extends into the area between the oILR and the iILR,
gradually diminishing in the counter-clockwise direction. The
transition from trailing to leading spiral arms occurs at a distance
of approximately $r\approx0.4$ from the center. In the case with
$c_s=10\,\rm{km\,s^{-1}}$, the prominent leading spiral arms appear
broader, emerging again close to the ends of the
major axis of the central, low-surface-density oval area.

We identify a spatial correlation between the clustering of orbits at the inner 
and outer edges of the ILR region in Fig.~\ref{fig:krun1_orbs} and the dense areas 
(depicted in yellow) that appear at approximately the same distances from the 
center, but at varying angles, in the gaseous response models of 
Fig.~\ref{fig:r1l2}a and Fig.~\ref{fig:r1l2}b. The orbital crowding in 
Fig.~\ref{fig:krun1_orbs} takes place at distances, $r$,
$058<r<0.75$ (outer, along the major axis) and $r\approx 0.32$ (inner, along the minor axis).
In Fig.~\ref{fig:r1l2}a the yellow part of the pseudo-ring reaches a distance $r\approx 
0.74$, while the outer tip of the leading spiral arms is at $r\approx 0.32$.  In 
Fig.~\ref{fig:r1l2}b the densest parts of the outer x2 region are located at 
$r\approx 0.72$ and the densest parts of the leading spiral arms are at 
$r\approx 0.3$.

In Fig.~\ref{fig:lineR1_10_300a}a, we consider a slit positioned on the snapshot 
depicted in Fig.~\ref{fig:r1l2}b. This slit extends from the upper left to the lower 
right side of the figure, intersecting all overdense regions and traversing 
the system's center. It has the form of a long, red arrow, with the beginning of 
the slit's line segment marked with a red heavy dot. Additionally, we have 
overplotted isodensity curves, explicitly showing that all surface density 
peaks are located away from the major axis of the bar. Specifically, they 
emerge twisted away from the y axis, in the clockwise direction for the 
pseudo-ring~-~trailing spiral structure at the outer edge of the x2 region, and 
in the counter-clockwise direction for the leading spiral arms.

\begin{figure*}
    \centering
	\includegraphics[width=2\columnwidth]{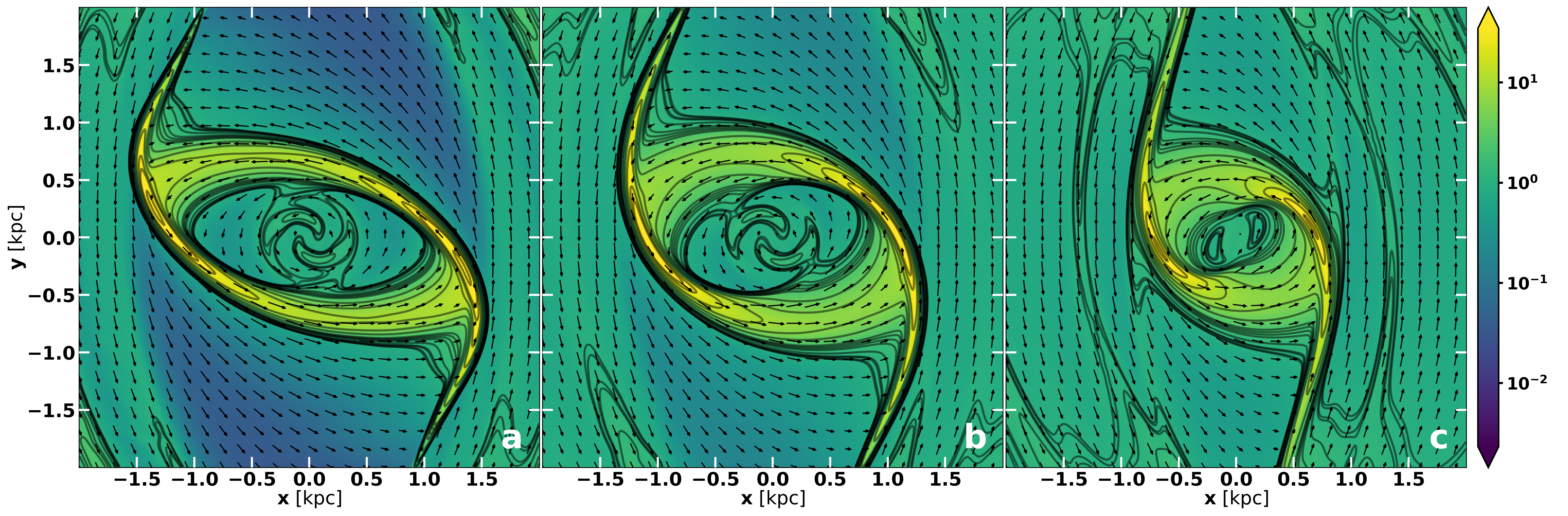}
    \caption{Model R001: Central regions of the three models during the growing 
phase of the bar. (a) $c_s=2$~km\,s$^{-1}$, (b) $c_s=10$~km\,s$^{-1}$, and
(c) $c_s=20$~km\,s$^{-1}$. The time of the snapshots corresponds to 1.7 pattern rotations
after the start of the simulation.}
    \label{fig:grow1}
\end{figure*}

In Fig.~\ref{fig:lineR1_10_300a}b, the blue curve represents the variation in gas 
surface density along the slit. Distances are measured either along the slit 
starting from its beginning (given on the lower side of the figure's box) or 
along the x axis (given on the upper side of the figure's box). The peaks of 
the curve, progressing from left to right, coincide successively with distinct 
features: the dust lane shock (a minor peak around $x\approx -1.02$), the 
trailing spiral (intercepted by the slit at a point not at its maximum surface 
density, approximately at $x\approx -0.45$), and the leading spiral arm (at 
$x\approx -0.2$). Beyond $x>0$, these local maxima reappear in a mirrored sequence.
In the central region, the blue curve has a clear minimum, indicating the presence of
the central oval area characterized by low surface density. The relatively 
empty region at the center of the model aligns with results from models with 
similar sound speeds by other authors \citep[e.g.,][]{at05}.

The orange curve in Fig.~\ref{fig:lineR1_10_300a}b represents the velocity 
components projected onto the slit, denoted by $u_{\parallel}$ (given in the 
right-hand side of the figure's box). The curve is anti-symmetric with respect 
to the center of the slit. Starting from the beginning of the slit and moving
from left to right, we first encounter a sharp drop in $u_{\parallel}$, corresponding to the minor 
peak of the surface density around $x\approx  -1.02$ (dust-lane feature), which
strongly suggests the presence of a shock. Then, as the slit enters the 
pseudo-ring region, $u_{\parallel}$ begins to rise, giving a $\Delta u_{\parallel} \approx 50$ 
close to the peak of the surface density of the trailing spiral along the slit around $x 
\approx -0.45$.

Subsequently, as we move along the slit toward the leading spiral arms, $u_{\parallel}$
drops and then has a local minimum at the local peak of the surface
density within the leading arm. We note a $\Delta u_{\parallel} \approx 100$ from
$u_{\parallel} \approx -50$ for $x \approx -0.2$, to $u_{\parallel} \approx 50$ for $x \approx
-0.1$. For $x < 0$, this change in $\Delta u_{\parallel}$ occurs as $u_{\parallel}$
transitions from negative to positive values. Within the range $-0.2 < x < 0.2$,
the curve displays a sinusoidal fluctuation, with $u_{\parallel}=0$ at the center of
the system. Beyond this region, the curve continues in a manner consistent with
its anti-symmetric nature.

In contrast to the morphologies depicted in Figs.~\ref{fig:r1l2}a,b, the case with 
$c_s=\rm{20\,km\,s^{-1}}$ lacks a discernible central leading spiral feature 
(Fig.~\ref{fig:r1l2}c). We can observe only a tendency for one of the innermost
isodensity contours to form a leading structure. Apart from the absence
of the leading spiral, the pseudo-ring structure appears shrinked, extending approximately to $x_{\text{max}} 
\approx 0.65$ in the x direction and to $y_{\text{max}} \approx 0.4$ in the 
y direction.

Noticeably, the crest of the pseudo-ring displays a set of two distinct local 
surface density peaks. This is visible in Fig.~\ref{fig:r1l2}c, where we have 
superimposed characteristic isodensity curves in the central area of the model. 
The set forms roughly along the major axis of the central, low density, oval, 
exhibiting two local surface density enhancements. The location and structure 
of these enhancements is reminiscent of the ``twin peaks'' feature encountered 
in the CO emission in the central region of barred galaxies such as NGC~3351 or 
NGC~6951 \citep{kjw92}. Additionally, it is worth noting that the pseudo-ring
appears to be inclined in the opposite direction, with respect to the minor axis of the system,
compared to the pseudo-rings observed in the $c_s=2$ and $10\,\rm{km\,s^{-1}}$ cases.
However, this pseudo-ring area is located within a green-blueish region of lower, though not 
zero, surface density, bordered by the two dust-lane shocks and extending up to 
$\vert y\vert \approx 1.5$. Due to the presence of this region of lower surface density, 
in Fig.~\ref{fig:r1l2}c the density outside the pseudo-ring area does not experience 
a sudden decline, in contrast to the cases depicted in Fig.~\ref{fig:r1l2}a,b.

By comparing the response models shown in Fig.~\ref{fig:r1l2}a,b,c, it becomes 
evident that pseudo-ring structures are consistently embedded within a “cloud” 
of low surface density (depicted in green-blue hues). This cloud encircles the 
pseudo-rings and extends toward the line segments due to the dust-lane shocks 
close to the x2 region, i.e., the region in which a nuclear ring-like structure 
is formed. Both the width of this region and its inclination relative to the 
bar's minor axis increase as the sound speed rises, as seen in the progression 
from Fig.~\ref{fig:r1l2}a to Fig.~\ref{fig:r1l2}c. We can say that while the cloud rotates clockwise, 
the pseudo-ring rotates counterclockwise and shrinks.

In Fig.~\ref{fig:r1l2}a, with $c_s=2\,\rm{km\,s^{-1}}$ the x2 region reaches a 
distance $\vert x\vert \approx 1.4$, slightly smaller than the extent of the largest x2 
periodic orbit of the model. In Fig.~\ref{fig:r1l2}b, with $c_s=10~\rm{km\,s^{-1}}$ 
the ring has shrunk, extending up to $\vert x\vert \approx 1$, while in the model with 
$c_s=20~\rm{km\,s^{-1}}$ (Fig.~\ref{fig:r1l2}c) the pseudo-ring structure is 
confined within a maximum $\vert x\vert \approx 0.65$.

To determine the formation time of the leading spiral feature, we investigated 
the initial phases of the model, specifically by studying its behavior during 
the growing phase. The response after 1.7 pattern rotations, for the three cases
with different sound speeds, is given in Fig.~\ref{fig:grow1}. A 
clearly discernible weak leading spiral emerges in the cases of $c_s=2$ and 
10~km\,s$^{-1}$ (Fig.~\ref{fig:grow1}a and Fig.~\ref{fig:grow1}b). In Fig.~\ref{fig:grow1}a 
the arms almost reach the ring, while in Fig.~\ref{fig:grow1}b the faint leading 
spiral is confined within a smaller radius. At any rate, for $c_s=20$~km\,s$^{-1}$
(Fig.~\ref{fig:grow1}c) no apparent leading spiral emerges in this region
during the growing phase. We only find two local density maxima squeezed at the 
very center of the system, which might be interpreted as reminiscent of a 
leading structure. Nonetheless, even this is a transient feature.

\begin{figure*}
    \centering
	\includegraphics[width=2\columnwidth]{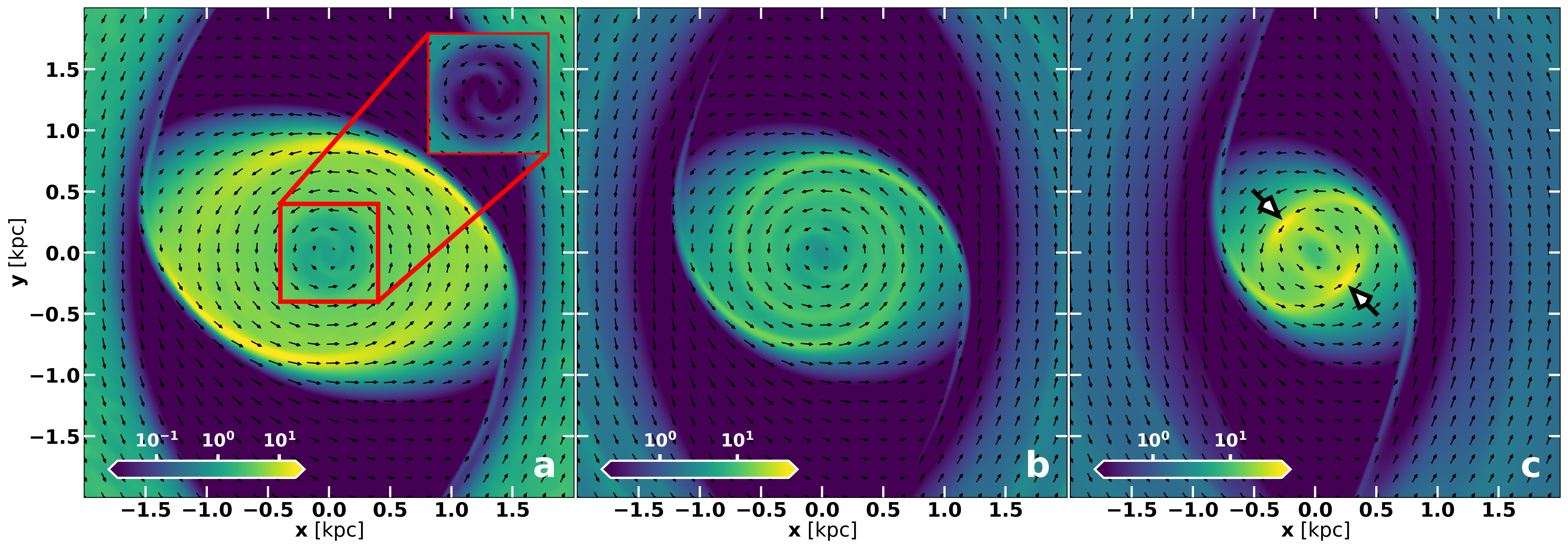}
    \caption{Model R010: Central regions of three models with different sound
speeds, (a) $c_s=2$~km\,s$^{-1}$, (b) $c_s=10$~km\,s$^{-1}$, and
(c) $c_s=20$~km\,s$^{-1}$. In (a) we give, in an embedded frame,
the faint leading spiral with higher contrast. As in 
Fig~\ref{fig:r1l2} all discussed features are embedded within an oval region, the 
inclination of which, with respect to the x axis, increases with sound speed.}
    \label{fig:r10}
\end{figure*}

The fact is that, in all cases, ultimately, following the conclusions drawn from
the growth phase, the response to the bar potential with full amplitude results in 
the development of the ring-like structures observed in Fig.~\ref{fig:r1l2} for each model.
The morphologies shown in Fig.~\ref{fig:r1l2} are already present during the growth phase
and become fully established approximately 1–2 dynamical times after the bar growth is complete.

Summarizing the main findings of our study of the R001 model, we can say that:
\begin{itemize}
 \item The emergence of the leading spiral feature is not exclusively dependent
on the underlying gravitational potential. In model R001 we find it for $c_s$=2 and 
10~km\,s$^{-1}$, but not for 20~km\,s$^{-1}$, or at least it is so weak in the latter case
that it becomes indiscernible in the gas surface density maps of our model. This already holds
during the growing bar amplitude period, which lasts for three pattern rotations.
 \item In all three cases of model R001 with different sound speeds there is 
no density maximum at the center of the system. An elliptical region, with 
a different inclination in each of the three cases, forms at the center and 
always has a lower surface density than its surroundings.
 \item While the morphology found in the cases with
$c_s=\rm{2\,km\,s^{-1}}$ and $c_s=\rm{10\,km\,s^{-1}}$ is rather
unusual, if observed at all, in the central regions of barred
galaxies, morphologies resembling that of the model with
$c_s=\rm{20\,km\,s^{-1}}$ are more commonly found. Examples include
galaxies such as NGC~1530, which exhibit
similar features as the model during the growing phase
(Fig.~\ref{fig:grow1}c), or NGC~1433, with a morphology 
similar to that of the model at the end of the simulation
when the full potential is considered (Fig.~\ref{fig:r1l2}c).
\end{itemize}

\subsection{Model R010 - A ``round'' bar}
\label{sec:R010}

\begin{figure}
    \centering
	\includegraphics[width=0.8\columnwidth]{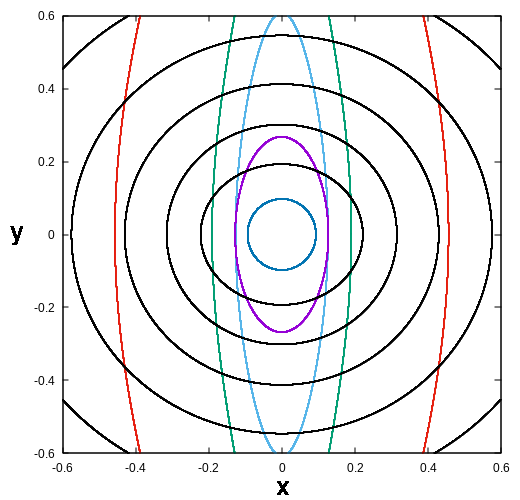}
    \caption{Model R010: x1 and x2 orbits in the central $1.2 \times 1.2$~kpc 
region. The x1 periodic orbits are given with colors, while the x2 in black. The almost 
circular shape of the orbits is associated with the trailing spirals appearing 
in the gas responses in Fig.~\ref{fig:r10}a,b.}
    \label{fig:krun10_orbs}
\end{figure}

Model R010 corresponds to RUN 10 in \citet{ath92a}. It is very similar to R001. 
The imposed bar is simply rounder ($a/b = 1.5$, against $a/b = 2.5$ in R001).
This alteration in the topology of the bar introduces several new 
characteristics in the central regions of the response models. Compared to model 
R001, we find discrepancies in the structure of the nuclear ring and the 
region near the center of the system. However, these variances primarily pertain 
to the cooler models with $c_s$=2 and 10~km\,s$^{-1}$, whereas for
$c_s$=20\,km\,s$^{-1}$, the deviation from R001 is minimal. The three R010 
models are illustrated in Fig.~\ref{fig:r10} (compare with the corresponding models 
for R001 in Fig.~\ref{fig:r1l2}).

A nuclear ring, or pseudo-ring, at the oILR region is less pronounced this time. 
In Fig.~\ref{fig:r10}a ($c_s$=2~km\,s$^{-1}$) we observe that the inclined, with 
respect to the minor axis of the bar, x2 oval is penetrated by the inward 
extension of the dust lane shocks, forming two arcs of high surface density  
along the periphery of the ILR region. These arcs seem to fade out quite 
abruptly as they tend to spiral inward in a trailing sense.

\begin{figure*}
    \centering
	\includegraphics[width=2\columnwidth]{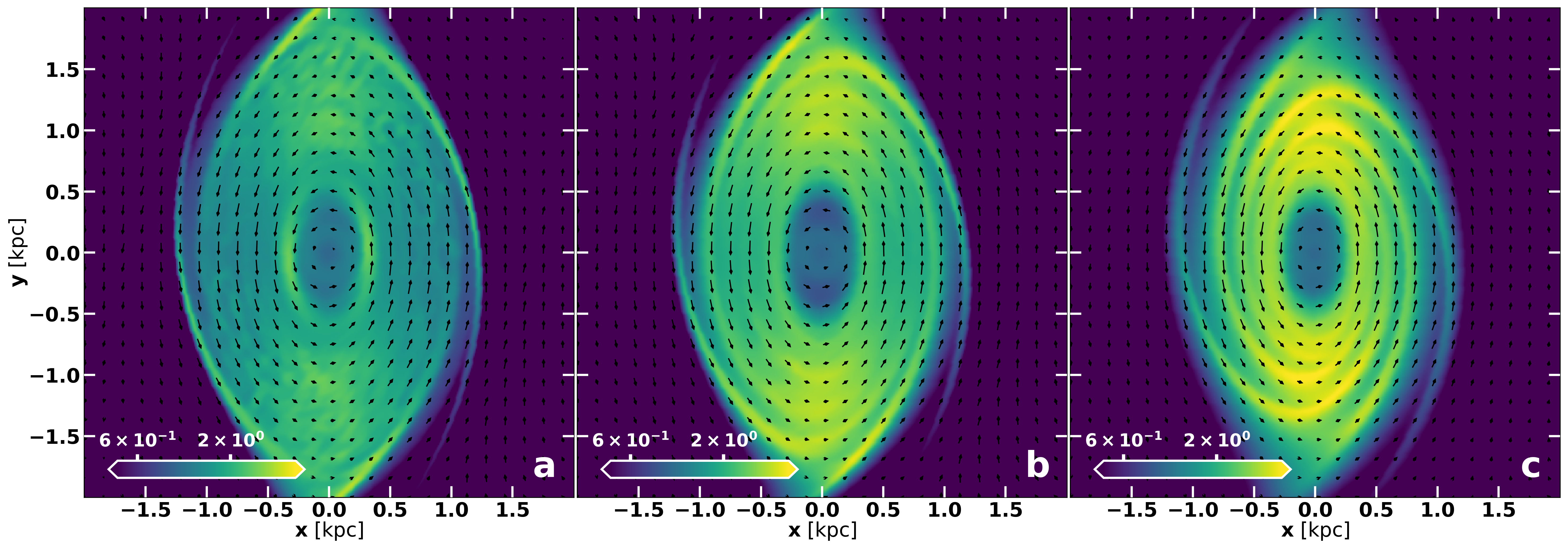}
    \caption{Model R153: Central regions of three models with different sound
speeds, (a) $c_s=2$~km\,s$^{-1}$, (b) $c_s=10$~km\,s$^{-1}$, and (c) $c_s=20$~km\,s$^{-1}$.
The responses are characterized by the presence of a multiarmed spiral structure, which
becomes more pronounced with higher sound speeds.}
    \label{fig:r153}
\end{figure*}
\begin{figure*}
    \centering
	\includegraphics[width=2\columnwidth]{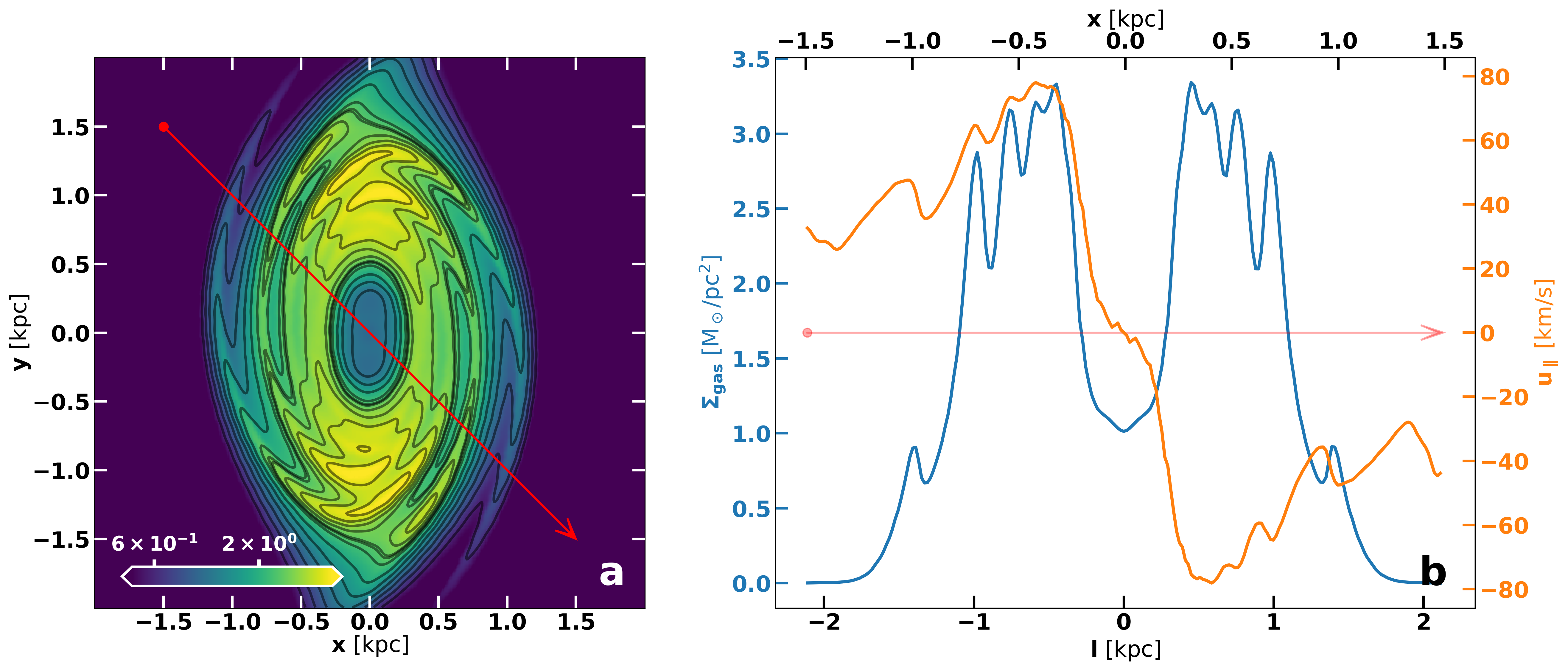}
    \caption{Model R153: Variation in gas surface density ($\Sigma$) and the 
velocity component along a defined slit ($u_{\parallel}$) for $c_s$=20~km\,s$^{-1}$.
Panel (a) shows the slit’s location, while panel (b) presents the variations, 
with the blue curve indicating $\Sigma$ and the orange curve representing 
$u_{\parallel}$. The labels on the axes are the same as in Fig.~\ref{fig:lineR1_10_300a}.}
    \label{fig:r153_20_slit}
\end{figure*}

For a larger $c_s$, i.e., for $c_s$=10~km\,s$^{-1}$ the overall morphology of the 
x2 region appears rounder. The shocks originating from the dust lanes continue
existing as higher gas density features, penetrating the ILR region and curving inward in a 
spiral manner toward the center (Fig.~\ref{fig:r10}b). We identify trailing spirals 
which reach a distance of $r \approx 200$~pc from the center. The 
arm-interarm contrast is low. It is hard to say if these spiral arms are 
logarithmic with a very small pitch angle, or Archimedean\footnote{The 
Archimedean spiral has the property that any ray from the origin intersects 
successive turnings of the spiral at points with a constant separation 
distance.}. This bisymmetric trailing spiral can be traced also in the 
$c_s$=2~km\,s$^{-1}$ R010 model. However, its very low amplitude makes it 
difficult to discern clearly by eye.

We note that tightly wound bisymmetric spirals of a similar morphology 
are encountered throughout the entire disk (beyond the ILR region) in stellar 
and gaseous models responding to weak spiral potentials  
\citep{p91,phgasLN}. In those cases the x1 periodic orbits supporting the spiral 
arms are almost circular. These works suggest that small disturbances to an 
axisymmetric background in spiral galaxy models cause nearly circular stable 
periodic orbits to be populated, offering the background support for the
emergence of tightly wound (of Sa type) stellar spiral arms and gaseous arms in
the corresponding hydrodynamic response models. We will see below that imposed
``round'' and weak perturbations are associated with tightly wound spirals
in the response models.
 
In the $c_s$=20~km\,s$^{-1}$ case (Fig.~\ref{fig:r10}c), the presence of trailing 
spirals in the x2 region persists, yet they are of a grand-design 
character and terminate suddenly, giving rise to 
two localized surface density peaks (indicated with heavy arrows). Also 
in this case we have a morphology reminiscent of ``twin peaks'' featured in
\citet{kjw92}. There is again a central low-density region, which 
takes on an oval shape, akin to the warm R001 model. Although the 
behavior of the warm R010 model bears a general resemblance to that of R001,  
the pseudo-ring structure, if present at all, is much weaker and less 
well defined (cf. Fig.~\ref{fig:r10}c and Fig.~\ref{fig:r1l2}c).

Upon careful examination of the central regions depicted in Fig.~\ref{fig:r10} for 
the three R010 models, it becomes apparent that the centers of these models 
exhibit lower densities compared to their surroundings. In Fig.~\ref{fig:r10}a (with 
$c_s$=2~km\,s$^{-1}$), a rather circular, disk-like region with a radius 
of approximately 400 pc is formed. The density within this region is notably 
lower than that of the surrounding x2 region, as indicated by the color bar at 
the bottom-left side of the frame. Within this disk, a leading spiral feature 
emerges, reaching an innermost distance from the center of about 8 pc. This 
feature is faint, and thus, for clarity, we also provide a frame with higher 
contrast embedded in Fig.~\ref{fig:r10}a.

In Fig.~\ref{fig:r10}b (with $c_s$=10~km\,s$^{-1}$), the dominant trailing spiral 
within the x2 region extends to an innermost distance from the center of
approximately 300 pc. At this distance, the winding direction of the 
bisymmetric spiral reverses, and a faint leading spiral is formed, also in this case (not 
readily visible in Fig.~\ref{fig:r10}b). The 
darker color of the region very close to the center once again indicates a lower 
central density compared to its surroundings.

Investigating the orbital dynamics of the system, we realize that
the area between the center of the system and the iILR is again dominated by 
order. There is no trace of chaos and the only families of periodic orbits 
existing in the region are x1, x2 and x3. 

However, the details of the shape of the orbits in the R010 case differ from 
those of R001. Except for the outermost x2 orbit at $E_J\approx -160000$, all
other x2 orbits are almost circular. The difference is evident in 
Fig.~\ref{fig:krun10_orbs}, which depicts the periodic orbits of the x1 and x2 
families at the central  $1.2 \times 1.2$~kpc (compare with the periodic orbits in the 
corresponding region of model R001 in Fig.~\ref{fig:krun1_orbs}). The almost 
circular shape of the x2 orbits is reflected in the formation of the tightly 
wound spirals in Fig.~\ref{fig:r10}b, with $c_s$=10~km\,s$^{-1}$. This 
spiral pattern is also present in Fig.~\ref{fig:r10}a, albeit with a significantly 
reduced amplitude, making it much less noticeable. However, this correspondence 
to the orbital background of the model is lost when we follow the response with 
$c_s=20$~km\,s$^{-1}$ (Fig.~\ref{fig:r10}c).
 
The main findings of our study of the R010 case are the following:
\begin{itemize}
\item Model R010 indicates that round bars with an axis ratio of $a/b = 1.5$
encounter challenges in generating nuclear rings at the periphery of the x2 
region akin to the one illustrated in Fig.~\ref{fig:r1l2} for R001.
\item There is a trend toward the emergence of trailing, tightly wound nuclear 
spirals within the x2 region of the model. In cases where these spirals form, 
they eventually evolve into a leading, bisymmetric spiral structure, which 
prevents them from extending into the centralmost region of the system.
\item The case with $c_s=\rm{20\,km\,s^{-1}}$ exhibits a response similar to 
that of the corresponding R001 case. However, we highlight the absence of a clearly 
defined nuclear ring.
\end{itemize}

\subsection{Model R153 - A bar without ``ILRs''}
\label{sec:R153}

Model R153 corresponds to RUN 153 in \citet{ath92a}. It has a very fast rotating 
bar, which brings $r_L$ at 3~kpc. Due to the high value of $\Omega_B = 
77.726$~km\,s$^{-1}$\,kpc$^{-1}$, the locations of the resonances are 
considerably displaced with respect to the former cases. The whole extent of the 
response bar, identified by the extent of the longest x1 periodic orbit, 
reaches a distance $y_{max} \approx 2.4~$kpc along its major axis. There is no 
ILR in this model. In the axisymmetric case, $\Omega_B$ is above the $\Omega - 
\kappa/2$ curve and the orbital analysis shows the absence of a x2-x3 loop in 
the characteristic diagram.

The gaseous responses of the R153 model for the three distinct sound speeds 
we examined are illustrated in Fig.~\ref{fig:r153}. The presented results are contained within a
$4\times4$~kpc box, covering nearly the entire bar region. In the absence of a 
x2-flow, we find a lack of rings or pseudo-rings in the system. Instead, the gaseous
responses are characterized by the emergence of a multiarmed spiral structure, which
becomes more pronounced with higher sound speeds. In
Fig.~\ref{fig:r153}c, with $c_s=20$~km\,s$^{-1}$, a conspicuous fourfold spiral
pattern  is noticeable. The arms of this spiral pattern reach the outer regions
of the bar. Meanwhile, the central area is marked by a low-density oval, which
remains unaffected by the multiarmed spiral. In Fig.~\ref{fig:r153}b
($c_s=10$~km\,s$^{-1}$), only the outer segments of the arms are
distinguishable, whereas in Fig.~\ref{fig:r153}a with $c_s=2$~km\,s$^{-1}$, only two
spiral arm tips, tangential to the bar's oval boundary, are visible. In this
latter case, within the bar region, spiral arms are scarcely found in the
model's bar area. It is worth mentioning that multi-spiral responses have been
previously identified by other authors, emerging from nested sequences of
streamlines within regions governed by low eccentricity x1 periodic orbits in
the underlying potential, as demonstrated by \citet[][ - see e.g., figure 10 in 
Sormani et al 2015b]{sbm15II, sbm15III}. However, in those instances, they 
are located outside the bar region, which is characterized by the presence of typical
dust-lane shocks. We find something similar with the Sormani models in the
overall response of our R010 model, over the whole bar and disk area.

The slit profile for the R153 model with $c_s=20$~km\,s$^{-1}$ is given in 
Fig.~\ref{fig:r153_20_slit}. In Fig.~\ref{fig:r153_20_slit}a, the slit is depicted as a 
long red arrow, with its starting point indicated by a prominent red dot. The 
variation in surface density along the slit is depicted by the blue curve in 
Fig.~\ref{fig:r153_20_slit}b. Peaks in the curve indicate points where the slit 
crosses local surface density maxima, as is evident from the positions of the 
isodensity curves shown in Fig.~\ref{fig:r153_20_slit}a. Clearly the surface density 
decreases at distances $\vert x\vert \lessapprox 350$~pc.

The orange curve depicted in Fig.~\ref{fig:r153_20_slit}b demonstrates that there 
are no significant velocity component jumps along the slit. The discrepancies 
observed in the $u_{\parallel}$ component when traversing the local density maxima 
within the spiral arm region are on the order of a few percent, while in the 
low-density central region, such variations are not observed at all.

\begin{figure}
    \centering
	\includegraphics[width=0.975\columnwidth]{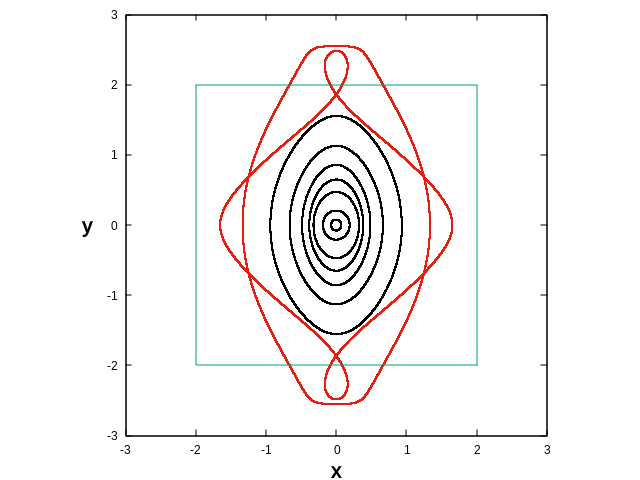}
    \caption{Model R153: Elliptical-like periodic orbits belonging to the x1 family, 
spanning from the system's center up to the 4:1 resonance, are depicted in 
black. Representatives of the x1 family at the 4:1 resonance region, as well as 
those from the 4:1 family, are illustrated in red. The light blue frame 
outlines the region of the gaseous responses presented in Fig.~\ref{fig:r153} and 
Fig.~\ref{fig:r153_20_slit}.}
    \label{fig:orbs153}
\end{figure}

Without x2 orbits present, we find only x1 periodic orbits with their major 
axis aligned along the major axis of the bar, extending to the center. These x1 
orbits exhibit the characteristic shape of the family, and in the central region 
($r \lesssim 200$), they become nearly circular (see Fig.~\ref{fig:orbs153}). 
Notably, there is almost no chaos in the central bar region of the model, as 
evidenced by examining successive surfaces of section at the corresponding $E_J$  
values.

\begin{figure*}
    \centering
	\includegraphics[width=2\columnwidth]{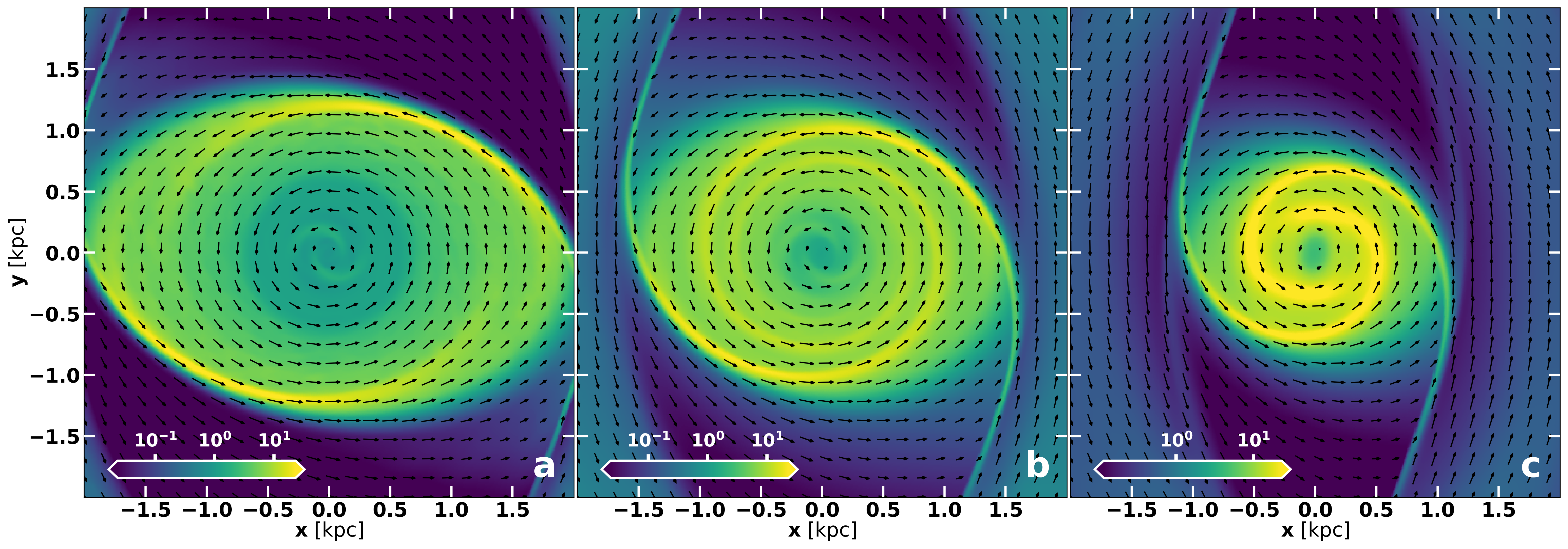}
    \caption{Model R192: Central regions of three models with different sound
speeds, (a) $c_s=2$~km\,s$^{-1}$, (b) $c_s=10$~km\,s$^{-1}$,
and (c) $c_s=20$~km\,s$^{-1}$. The model is characterized by the 
presence of trailing spirals in the x2 region, becoming more open with increasing
sound speed.}
    \label{fig:gas192}
\end{figure*}

Model R153 exhibits a type-I gap at the 4:1 resonance \citep{gco88}. In this 
type of gap, the characteristic of the bifurcated 4:1 family lies below the x1 
characteristic, in contrast to the relative positioning of the 4:1 
characteristic compared to the x1 characteristic in all other models, where it 
is located above x1 (type-II gap) \citep[see, for instance, model R001 
in][figure 2]{ath92a}. Consequently, as we approach the 4:1 resonance, we 
encounter periodic orbits typical for the 4:1 resonance region. The x1 orbits 
become diamond-like with loops, while the orbits of the 4:1 family are boxy, in 
alignment with the findings of \citet{gco88}. Representatives of both families 
are plotted with red color in Fig.~\ref{fig:orbs153}.
 
The main result of our ``no-ILR'' model is the absence of all morphological 
features in the central region resembling those we have discussed so far.
We only find, as in most other models in the present work, a centrally located
low-surface-density region.

The results described in this section for the response of the R153 model remain 
nearly identical for $\rho_c = 1.6$ (RUN 142 in the table of models in 
\citealt{ath92a}). This suggests that the resulting response is largely driven by 
the rapid rotation of the bar.

The appearance of multiarmed spirals identified within the main bar is particularly
interesting. This topic will be addressed in a forthcoming paper.

\subsection{Model R192 - A weak bar}
\label{sec:R192}

Model R192 represents a weak bar model (refer to Table~\ref{tab:mparams}). It 
corresponds to the case described as RUN 192 in \citet[]{ath92a}. The 
distinction from the reference case R001 lies in the $Q_m$ parameter, which is 
set to 1.5 instead of 4.5, while the other parameters remain as in R001 and 
R010.

Fig.~\ref{fig:gas192} illustrates the response in the central region of model R192 
for the three different sound speeds, as in the previous cases. Assuming a 
weaker bar results in a response more akin to the rounder bar of R010 than to 
the fiducial R001 case. Specifically, a leading spiral feature appears for  
$c_s=2$ and 10~km\,s$^{-1}$, but it is faint and barely noticeable. No such 
leading spiral feature is identified in the gaseous response for 
$c_s=20$~km\,s$^{-1}$. The low-surface-density region in the central area of the 
models persists across all three cases. In Fig.~\ref{fig:gas192}c this region 
shrinks within a radius of approximately 200 pc. As with the rounder bar model,
the weak bar one also shows a resemblance to the responses to weak spiral 
potentials presented by \citet{p91} and \citet{phgasLN}.

The case with $c_s=20$~km\,s$^{-1}$ for the weak bar model R192 exhibits the 
longest azimuthally extended grand design trailing spiral that penetrates the x2 
region, among the models studied. We emphasize the similarities between the responses 
to weak and round bar perturbations, as evident from the comparison between 
models R010 and R192.

\subsection{Model R901 - A high $\rho_c$ model}
\label{sec:R901}

\begin{figure}[h]
    \centering
	\includegraphics[width=\columnwidth]{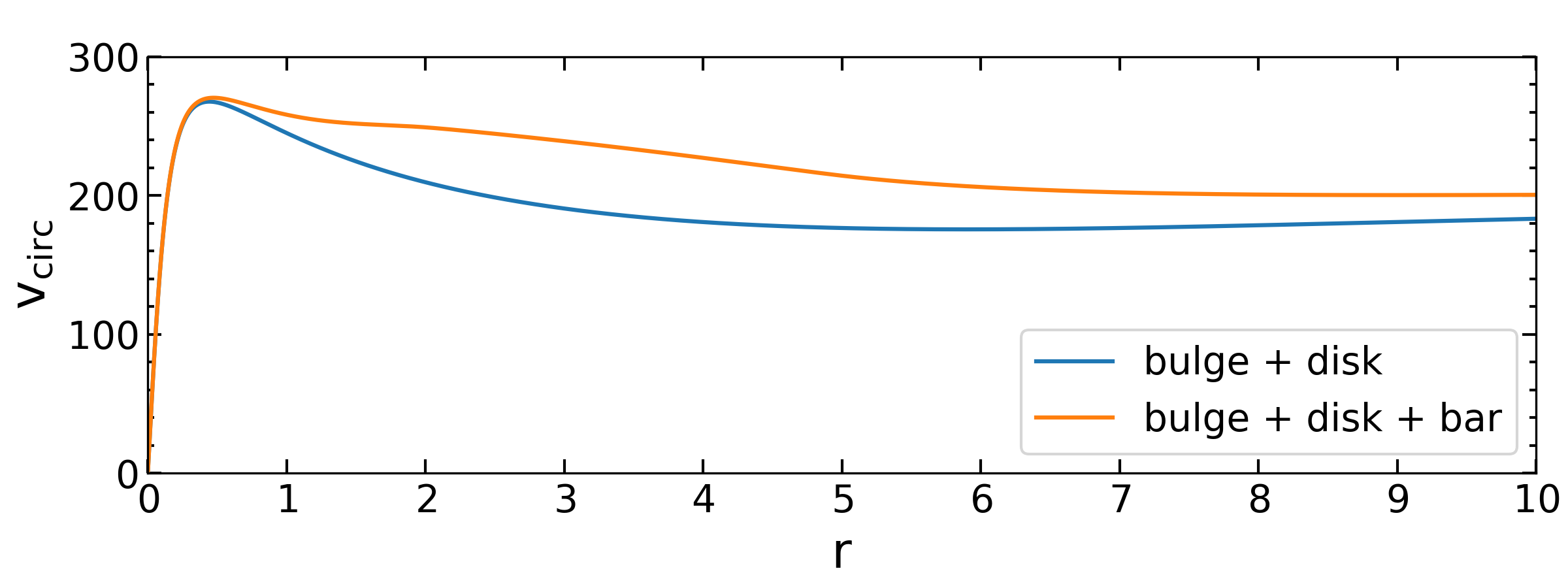}
    \caption{Rotation curve of model R901. The blue curve corresponds to the 
axisymmetric part of the potential, while the orange one to the full potential.}
    \label{fig:rc_krun901}
\end{figure}

In our study, we also examined a model with parameters that were not considered 
in \citet[]{ath92a} (last case in Table~\ref{tab:mparams}). This model, R901, 
has a high central concentration ($\rho_c$ is 8.3 times larger than in R001) 
that is reflected in the shape of the rotation curve (Fig.~\ref{fig:rc_krun901}). The 
peak of the rotation curve for R901 has been shifted inward compared to all 
preceding models, now residing well within the innermost kiloparsec region, with a 
maximum velocity of approximately 270~km\,s$^{-1}$.

\begin{figure*}
    \centering
	\includegraphics[width=2\columnwidth]{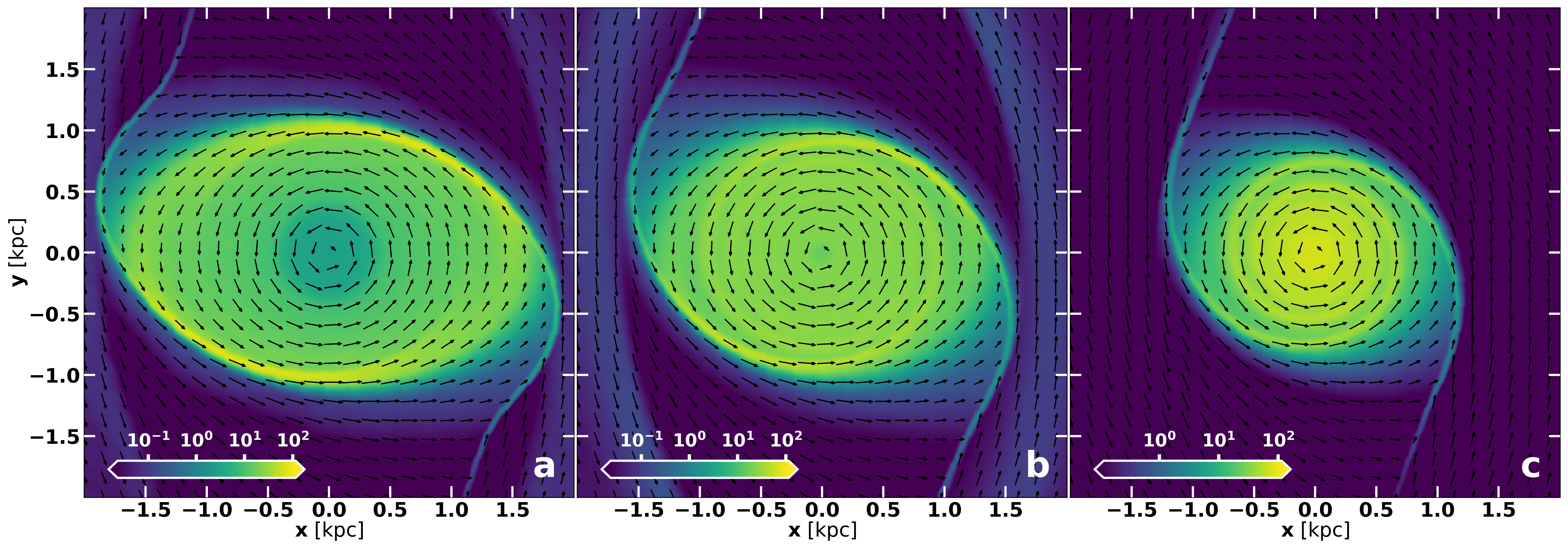}
    \caption{Model R901: Central regions of three models with different sound
speeds, (a) $c_s=2$~km\,s$^{-1}$, (b) $c_s=10$~km\,s$^{-1}$,
and (c) $c_s=20$~km\,s$^{-1}$. We find that, contrarily to all 
previous models, in (c) the surface density increases toward the center.}
    \label{gas901}
\end{figure*}
\begin{figure*}
    \centering
	\includegraphics[width=2\columnwidth]{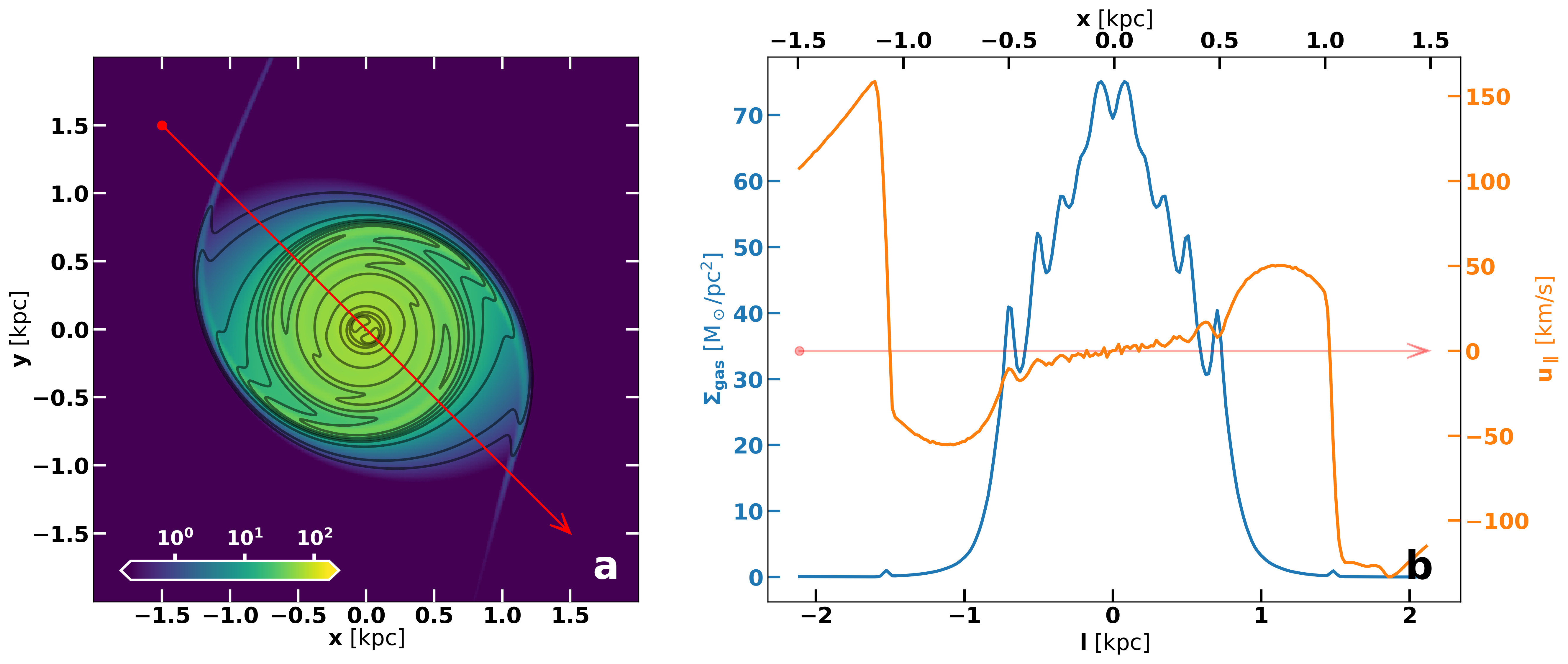}
    \caption{Model R901: Variation in gas surface density ($\Sigma$) and the 
velocity component along a defined slit ($u_{\parallel}$) for $c_s$=20~km\,s$^{-1}$.
Panel (a) shows the slit’s location, while panel (b) presents the variations, 
with the blue curve indicating $\Sigma$ and the orange curve representing 
$u_{\parallel}$. The labels on the axes are the same as in Fig.~\ref{fig:lineR1_10_300a}.}
    \label{lineR901_20_300a}
\end{figure*}

The response of the R901 model for the three different sound speeds in the 
central $4\times4$~kpc box is given in Fig.~\ref{gas901}. In this case, there 
are notable differences worth highlighting among the three cases. In 
Fig.~\ref{gas901}a, with $c_s=2$~km\,s$^{-1}$, at the outer border of the x2 
region we observe the formation of a pseudo-ring. Furthermore, the progression 
of the dust-lane shocks from the primary body of the bar inward to the x2 region 
is marked by a distinct folding pattern of these shocks, preceding their 
convergence with the pseudo-ring. In Fig.~\ref{gas901}b, with 
$c_s=10$~km\,s$^{-1}$, the pseudo-ring structure is less evident; 
nevertheless, we can observe the emergence of a relatively circular inner disk. 
This disk is characterized by the formation of a faint inner trailing spiral 
winding toward the system's center, as one can realize by a careful inspection 
of the figure. The outermost part of this spiral, at $r\approx 0.8$~kpc, at 
the border of the x2 region, has the highest surface density along the spiral,
similar to the corresponding case in R192 (Fig.~\ref{fig:gas192}).
The difference is that in the R901 case, the inner spiral is barely discernible.
Finally, in Fig.~\ref{gas901}c, with $c_s=20$~km\,s$^{-1}$, the dust-lane shocks
persist inward, forming this time an inner disk with a discernible, tightly wound spiral pattern.

However, the most noticeable distinction lies in the center of each of the three 
models. In Fig.~\ref{gas901}a, there is a prominent central low-density 
disk-like region with a radius of approximately $r\lesssim 350$ pc. In 
Fig.~\ref{gas901}b, the density decreases to a lesser extent, with the color 
coding indicating a density minimum only at the very center. The surface density 
in the x2 region appears much more uniform compared to Fig.~\ref{gas901}a. 
Finally, contrarily to the first two, and all previous models, in 
Fig.~\ref{gas901}c the surface density seems to increase toward the center. In 
order to have a closer look at this model we consider again the isodensities in 
the central region (Fig.~\ref{lineR901_20_300a}a) and the variation in the 
surface density and the velocity component along a slit (Fig.~\ref{lineR901_20_300a}b).

In Fig.~\ref{lineR901_20_300a}b, we note that the blue curve, representing the 
surface density variation along the slit shown in Fig.~\ref{lineR901_20_300a}a, 
rises toward the system's center, exhibiting fluctuations corresponding to the 
arms of the trailing spiral observed in the x2 region. Within the innermost 45 
pc, the gas density decreases by 7.8\%. No evident leading spiral is observed in 
this case. However, in Fig.~\ref{lineR901_20_300a}a, we can discern two density 
peaks, corresponding to the two local maxima of the blue curve near the center 
in Fig.~\ref{lineR901_20_300a}b. Their location and structure has, as in the 
R001 and R010 ``warm'' models (Figs.~\ref{fig:r1l2}c and \ref{fig:r10}c, respectively),
a ``twin peaks'' character. However, in this case, in the absence of a pseudo-ring,
they appear much closer to the center, where gas is accumulated.
In \citet{kjw92} an example with ``twin peaks'' close to the center
is given for the weakly barred galaxy (of SAB(rs)cd type) M101 (see their figures 1 and 3).
The variation in the orange curve indicates that practically the only shocks happen at the 
loci of the dust-lane shocks outside the x2 region. We conclude that the dust-lane shocks 
transport gas into the already gas-rich central region, where it is further 
compressed into smaller radii.

\begin{figure}
    \centering
	\includegraphics[width=\columnwidth]{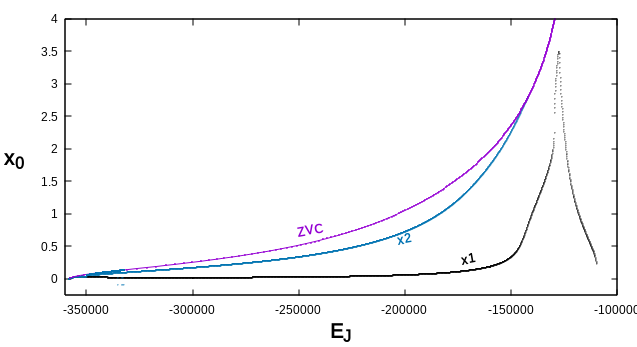}
    \caption{Model R901: Characteristics of the main families of the model. The zero 
velocity curve is indicated with ``ZVC'' and is given in purple color.}
    \label{fig:char901a}
\end{figure}

\begin{figure}
    \centering
	\includegraphics[width=0.8\columnwidth]{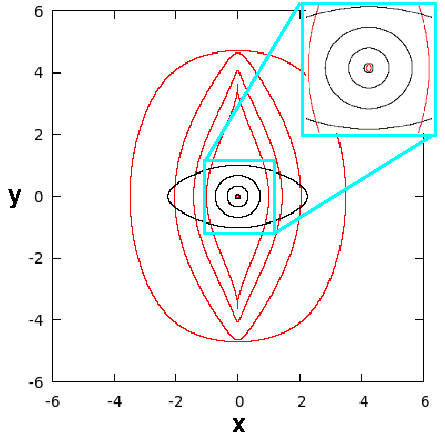}
    \caption{Model R901: Orbital backbone of the whole bar. In the embedded frame we zoom in the central region. The x1 and x2 orbits are depicted in red and black, respectively.}
    \label{fig:orbs901}
\end{figure}

To explore a potential connection between the orbital structure and the gas 
response, it is helpful to additionally analyze the model's orbital dynamics. The 
characteristic of the model is illustrated in Fig.~\ref{fig:char901a}. Both the x1 
and x2 families coexist within the energy range $-350000\lesssim E_J \lesssim 
-150000$. The stellar bar consists of orbits trapped around stable periodic 
orbits within this energy range. The local maximum of the x1 characteristic is 
at $E_j\approx -127000$. Spatially, the orbital backbone of the periodic orbits 
that support the bar of the model is given in Fig.~\ref{fig:orbs901}. In close
proximity to the center, for $E_J \lesssim -350000$, only almost circular
x1 orbits exist, with $r\lessapprox 30$~pc. Almost circular orbits belonging to both 
x2 and x1 family, not intersecting other periodic orbits, are present for 
$r\lessapprox 300$~pc, as can be inferred by the inspection of the
embedded zoom of the central region in the upper right corner of
Fig.~\ref{fig:orbs901}. This region corresponds to the central disk of low density
observed in Fig.~\ref{gas901}a.

The R901 model, with a sound speed of $c_s=$20~km\,s$^{-1}$, is the sole model 
encountered thus far in our study that, in practical terms, draws the gas close 
to the system's center. Consequently, we next examine two models with very 
weak bars to gain a deeper understanding of the role of the axisymmetric 
background, compared with the role of the bar component, in shaping the central 
1 kpc morphology.

\section{Comparison with extremely weak bars}
\label{sec:axisym}

Explanations for the presence of the leading spirals, as well as for the overall 
morphologies appearing in the gas in the central 1~kpc region, are sought within 
the framework of the linear approximation \citep[see, for 
example,][]{w94,wm04a,sbm15II}. However, such an approach is significantly 
different from that in which a firmly established bar is present. We aim to 
identify which characteristics of the models already emerge when gas  responds 
to very weak barred perturbations and which require stronger perturbations to 
manifest. In our models, we initially have a nearly axisymmetric phase during 
the early stages of our simulations, but this phase is transient and occurs only 
during the growth phase. Thus, we investigated the response to very weak 
barred potentials over several pattern rotations.

\begin{table}[h]
\caption{Dynamical parameters of the two extremely weak bar models.}
\label{tab:tab2}
\centering
\begin{tabular}{c c c c r c}
\hline\hline\\ [-1.5ex]
Model & $a/b$ &  $r_L$ & $Q_m/10^4$ & $\rho_c/10^4$ & $\Omega_B$ \\
\hline\\ [-1.5ex]
   R801 & 2.5 & 6.0 & 0.1 & 2.4  & 32.86\\
   R991 & 2.5 & 6.0 & 0.1 & 20.0 & 33.38\\
\hline
\end{tabular}
\end{table}

We present two extremely weak bar models with sound speed
$c_s$=10~km\,s$^{-1}$, the parameters of which are summarized in 
Table~\ref{tab:tab2}. Upon comparing the parameters of the model labeled R801
with those of R001 in Table~\ref{tab:mparams}, we note a difference in $Q_m$,
which has a value of 0.1 in the case of R801 instead of 4.5 for R001.
On the other hand, R991 shares similarities with R901, featuring a high $\rho_c$,
but differing in $Q_m$, which is once again 0.1 instead of 4.5. This practically
makes the bulge the important component of the models. The rotation curve of R801
is close to the one of R001 and that of R991 close to that of R901 and all
four models have two ILRs.

\begin{figure*}
    \centering
	\includegraphics[width=2\columnwidth]{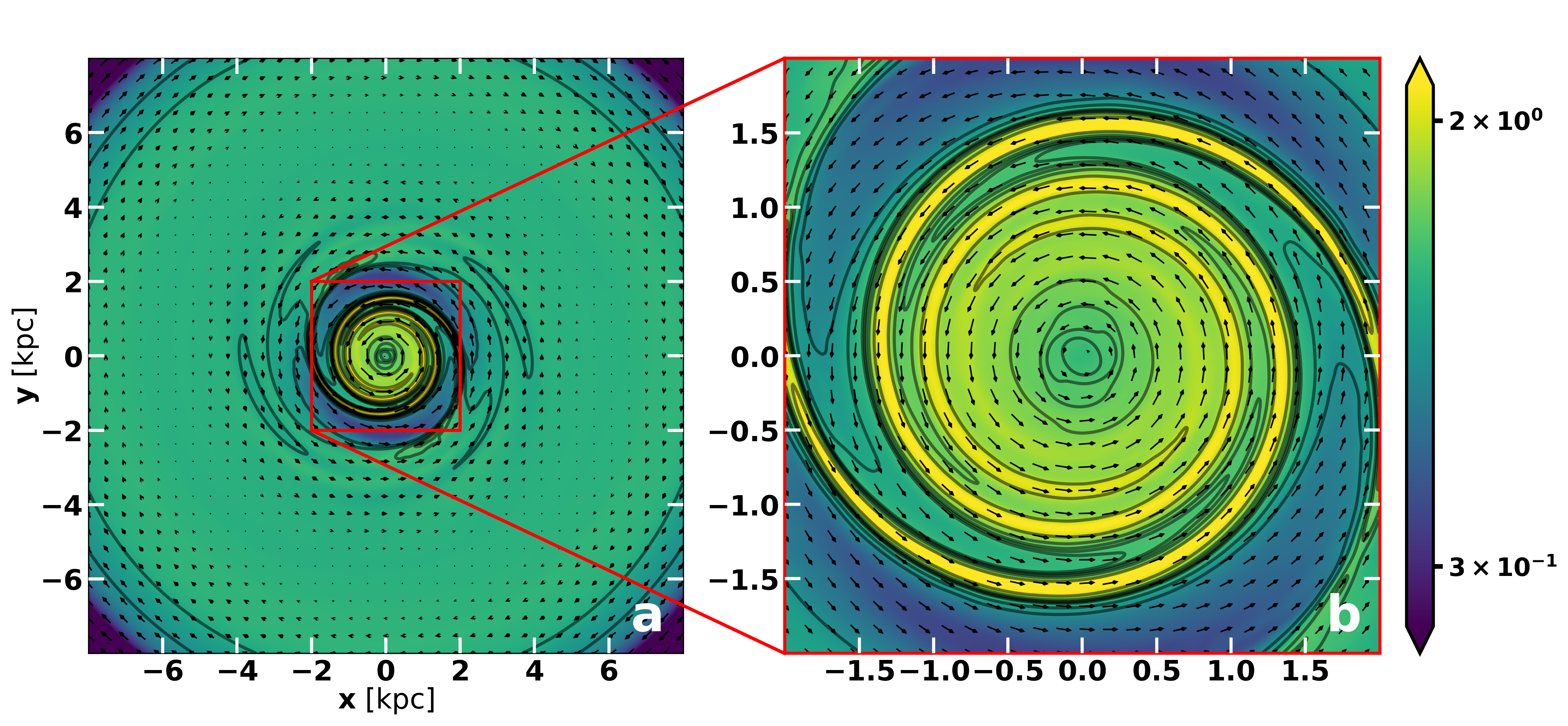}
    \caption{Model R801: Gas response of the model with $c_s$=10~km\,s$^{-1}$ in
(a) $16 \times 16$~kpc frame  and (b) in the central $4 \times 4$~kpc region. 
Apparently, the innermost 0.025~kpc have a lower surface density (greenish) compared to
the yellowish surroundings and the yellow spiral arms.}
    \label{aax801}
\end{figure*}

In Fig.~\ref{aax801} we give the overall R801 response (Fig.~\ref{aax801}a) and 
the response in the central $4\times4$~kpc box (Fig.~\ref{aax801}b). In 
Fig.~\ref{aax801}a, we note the absence of a discernible bar component and the 
absence of straight-line dust-lane shocks. The figures include the velocity vectors,
with the corotation region becoming apparent when considering their size and direction.
Within a radius of $r<2$ kpc, a bisymmetric spiral pattern emerges
that can be better appreciated in Fig.~\ref{aax801}b. It reaches an innermost radius of 
$r\approx 0.8$~kpc. The model is characterized by a  central oval area with 
lower density than its surroundings, as in the R001 model, this time for 
$r\lessapprox 0.45$~kpc. However, there is no apparent leading spiral feature 
formed in R801.

\begin{figure}
    \centering
	\includegraphics[width=\columnwidth]{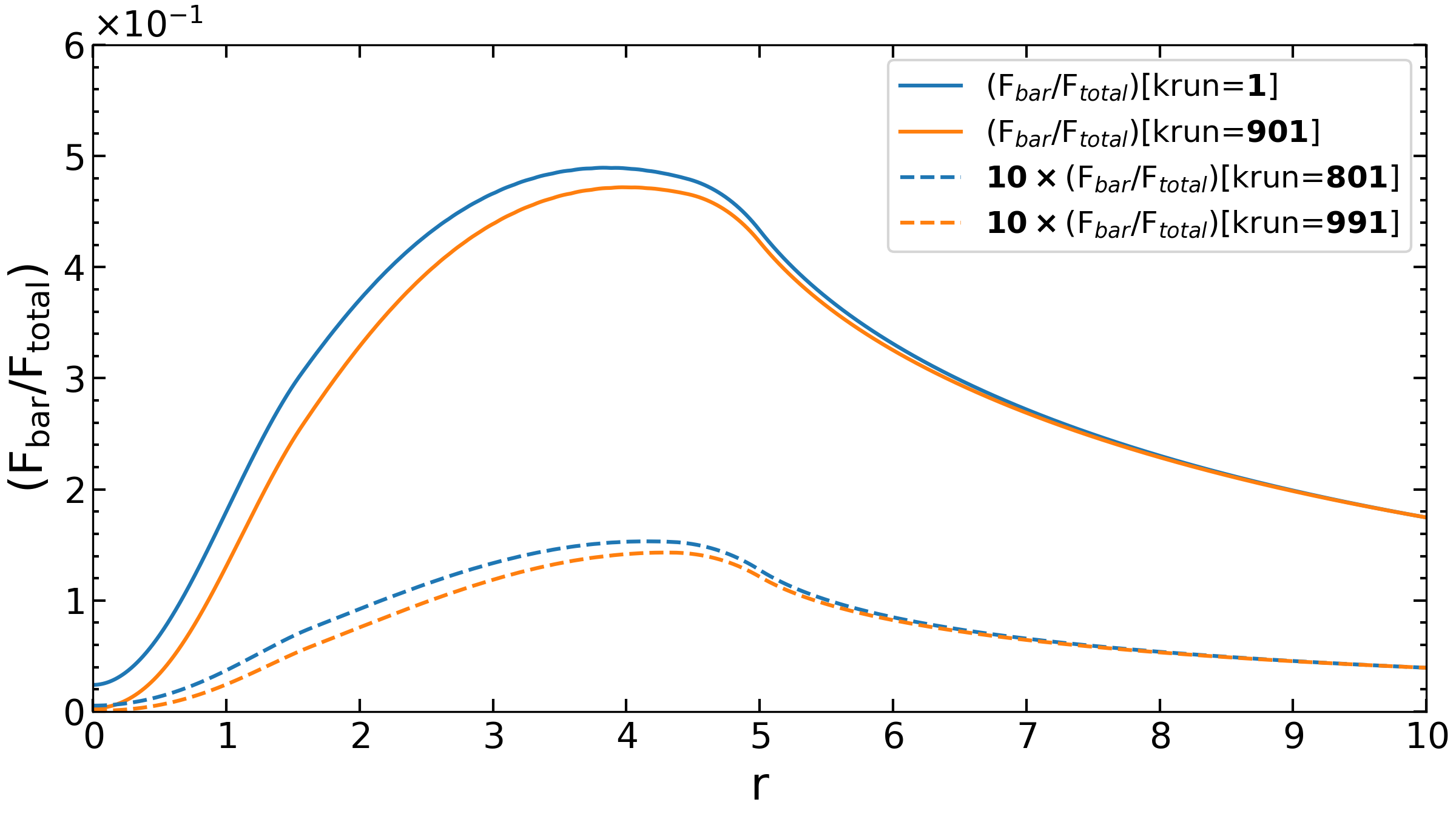}
    \caption{Perturbing force variations for the four models R001, R801, R901, and R991.}
    \label{fcentralbin}
\end{figure}

The lack of distinct dust-lane shocks beyond the region with $r\gtrsim 
2$ in R801 can be attributed to the much lower relative bar forcing, 
$F_{pert}=F_{bar}/F_{total}$, in this model with respect to that of R001. 
In Fig.~\ref{fcentralbin}, we present $F_{pert}$ for the two models discussed in 
this section (R801, R991), along with their corresponding models that include 
the full bar perturbations (R001 and R901, respectively). We find that for 
$2<r<6$, we have  $0.008<F_{pert}<0.015$ in R801 and $0.35<F_{pert}<0.48$ for R001.
This difference is also evident in the orbital structures: the weaker perturbation
in R801 results in nearly circular x1 orbits, while in R001, the x1 family’s
periodic orbits are elliptical-like.

In the central 1~kpc region, both R801 and R001 models exhibit low relative 
forcing; however, it is significantly lower in R801 than in R001. Specifically, 
the relative forcing for R801 is $F_{pert}<0.003$, while for R001 it is $F_{pert}<0.17$.
We highlight that both models have two ILRs (similar variation in their $\Omega(r) - 
\kappa(r)/2$ curves) and similar initial surface densities throughout their disks. 
Both models show a kind of trailing bisymmetric spiral in the 1~kpc region (more 
pronounced in the R801 case), but only R001 includes a clear leading spiral. 
The shapes of the three innermost isodensities in Fig.~\ref{aax801}b suggest a 
tendency toward the development of a leading component; however, this remains 
far from forming a well-defined leading spiral feature. Since the rest of the 
dynamical and hydrodynamical parameters are similar, the presence of the leading 
spiral feature (cf. Fig.~\ref{fig:r1l2}b and Fig.~\ref{aax801}b) is primarily 
influenced by the relative forcing, which is reflected in the shape of the
periodic orbits in the region.
 
\begin{figure*}
    \centering
	\includegraphics[width=2\columnwidth]{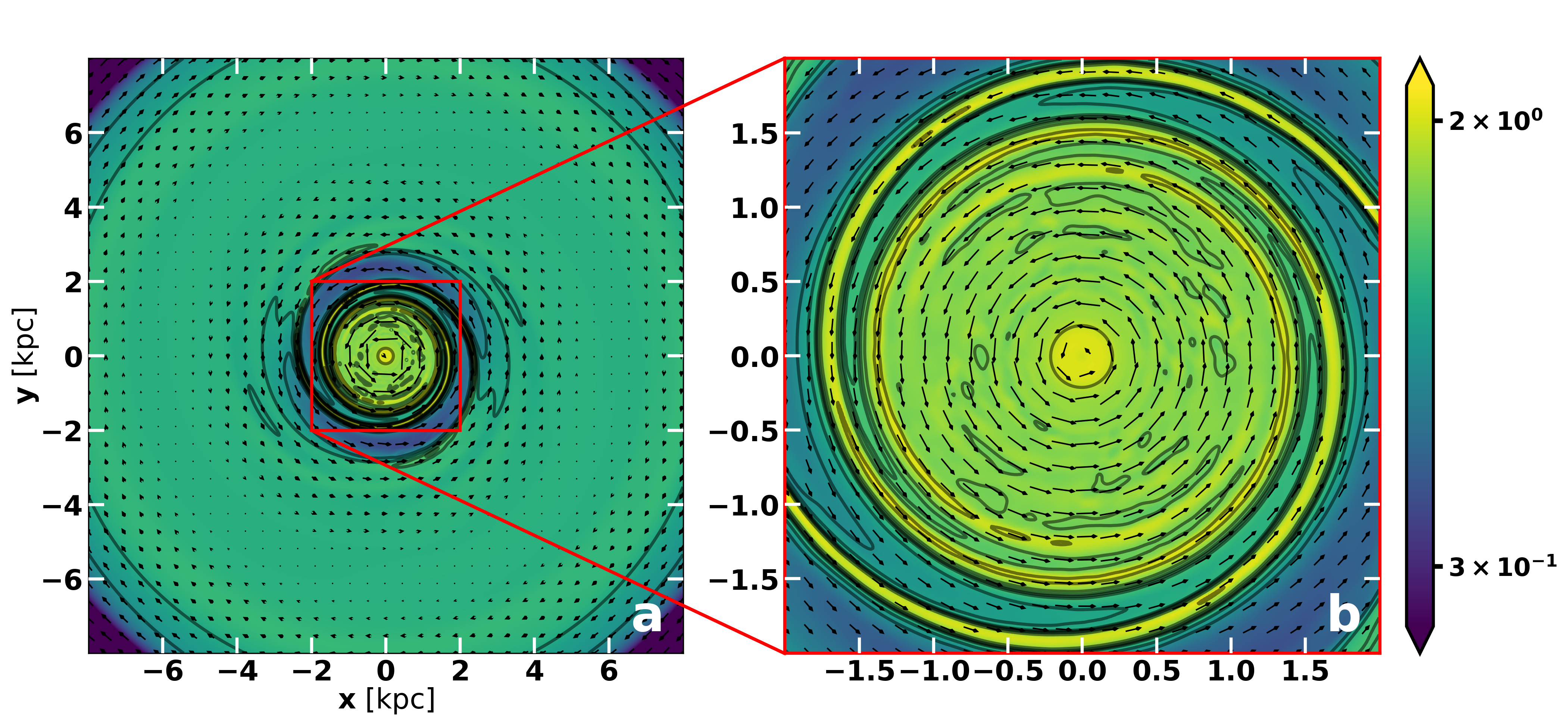}
    \caption{Model R991: Gas response of the model with $c_s$=10~km\,s$^{-1}$ 
in (a) $16 \times 16$~kpc frame  and (b) in the central $4 \times 4$~kpc region. 
The innermost 0.025~kpc have a higher surface density (yellow) compared to the
surrounding region (greenish).
}
    \label{aax991}
\end{figure*}

\begin{figure}
    \centering
	\includegraphics[width=0.8\columnwidth]{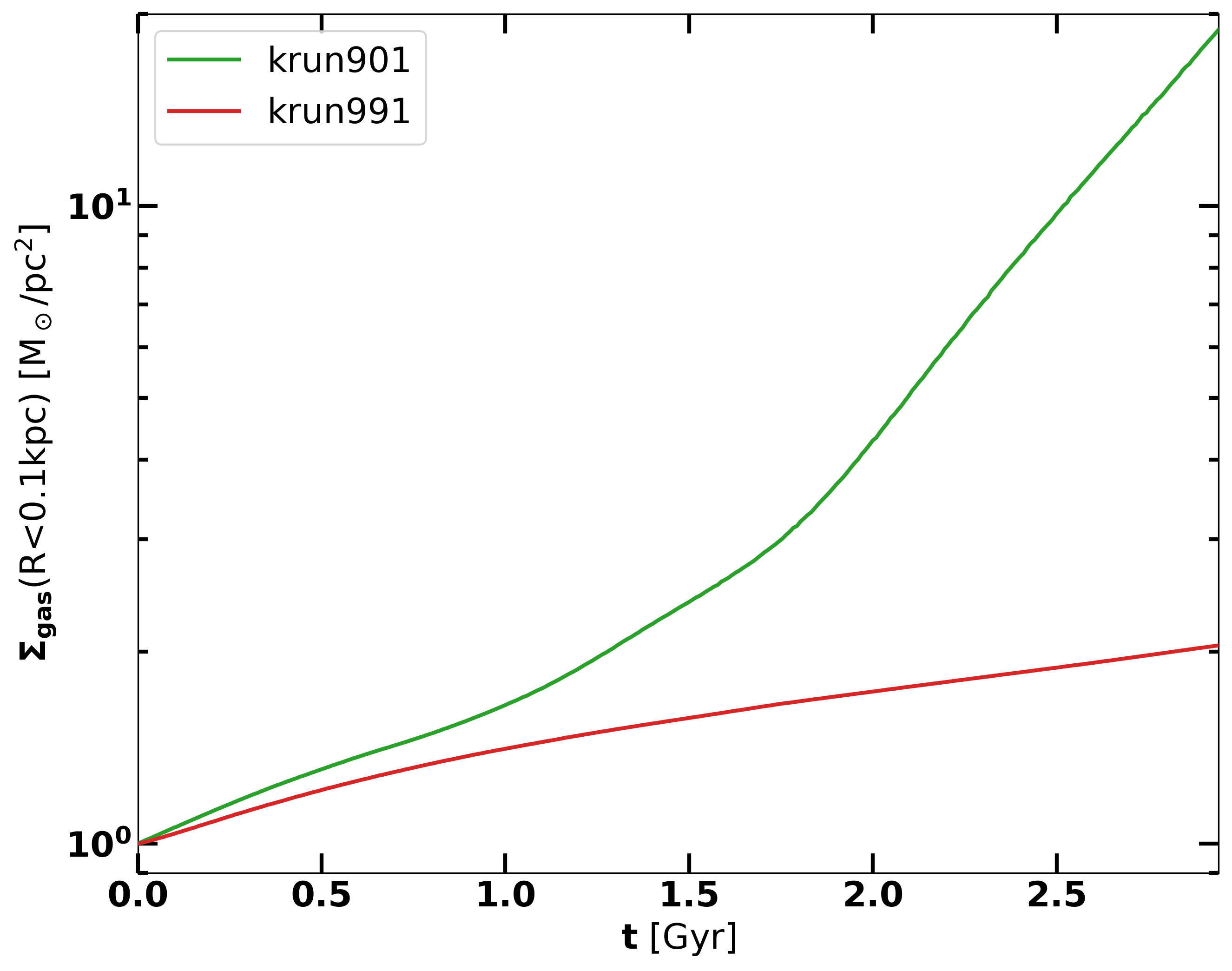}
    \caption{Surface density evolution in the central $r<0.1$~kpc region for 
models R901 and R991. We find that the R901 model accumulates an order of 
magnitude more gas in the central region than the almost axisymmetric model 
R991.}
    \label{sdcentralbin}
\end{figure}

The main distinction between the high $\rho_c$ models R991 and R901 primarily 
also lies in the absence of dust-lane shocks in the main body of the bar in R991,
which is nearly axisymmetric. As the center is approached in 
model R991, a tightly wound, trailing, bisymmetric spiral forms and becomes the 
dominant structure in the $2.5\lessapprox r \lessapprox 1.3$~kpc region 
(Fig.~\ref{aax991}a). In the corresponding region of R901, dust lane shocks are 
present, extending up to a nuclear pseudo-ring structure. Closer to the 
center ($r<1$~kpc), both models exhibit similarities, including a disk with 
nearly uniform density and weak spiral features. In R991, the spirals are 
fragmented, and the central surface density within $r<0.1$~kpc is higher 
compared to its surroundings (Fig.~\ref{aax991}b). In contrast, for the strong 
bar model R901, the surface density in this region is comparable to, if not 
lower than, that of its surroundings (Fig.~\ref{gas901}b).
However, in absolute terms the strong bar model R901 with 
$c_s$=10~km\,s$^{-1}$ accumulates considerably more gas in the central 
$r<0.1$~kpc region compared to 
R991, as was expected (Fig.~\ref{sdcentralbin}). 


\begin{figure*}
    \centering
	\includegraphics[width=2\columnwidth]{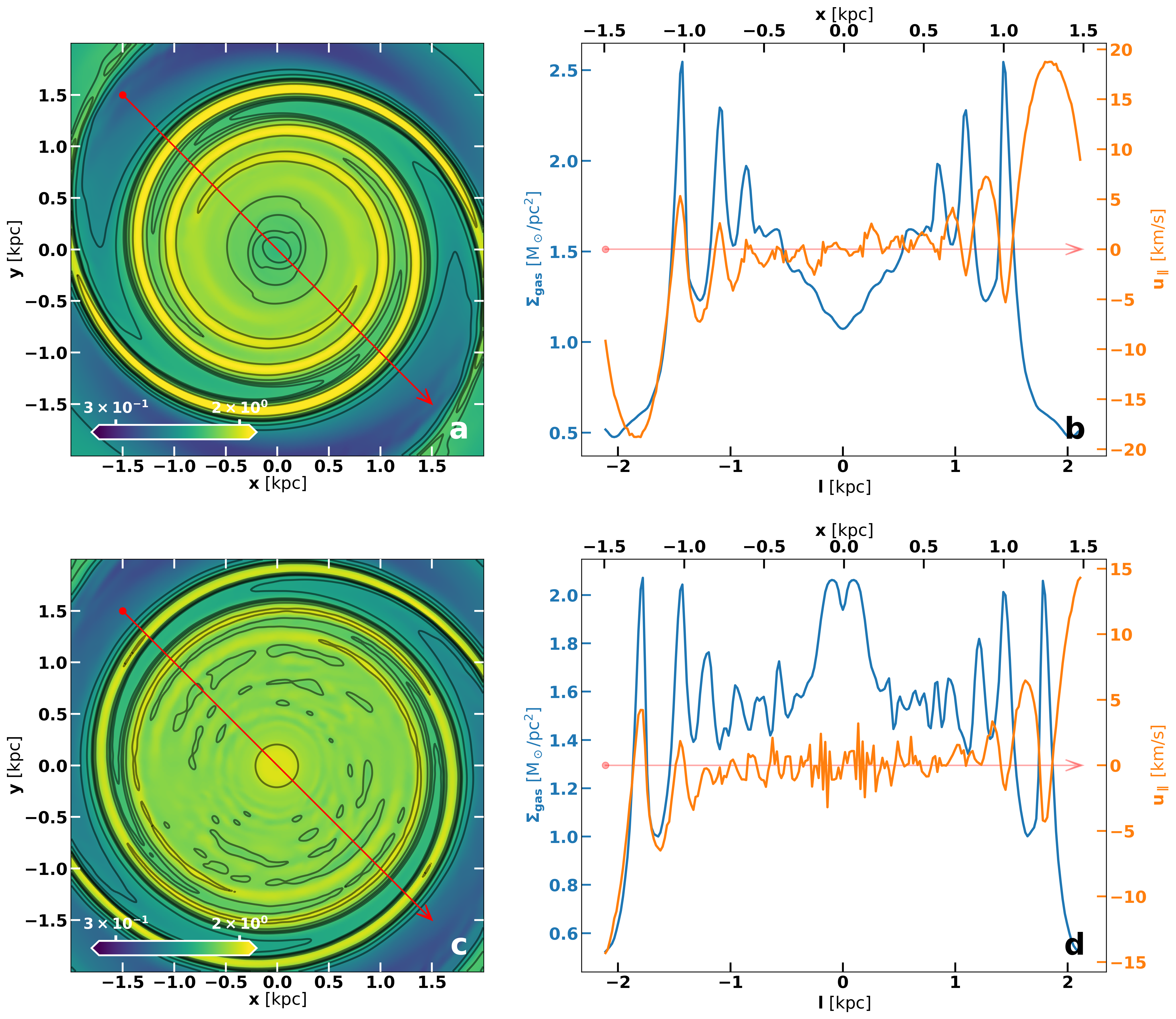}
    \caption{(a) and (b): Variation in gas surface density ($\Sigma$) and the 
velocity component along a defined slit ($u_{\parallel}$) for model R801 with
$c_s$=10~km\,s$^{-1}$. (c) and (d) Same figures for model R991. We find that if a 
high surface density is initially set at the center (R991), it persists throughout
the simulation; without it (R801), the central density remains low, 
lower than that of its surroundings, as in the case with the full potential.}
    \label{aaxlines}
\end{figure*}

To gain a clearer understanding of the central spiral patterns that develop in 
the two very weak bar cases, we examined the variations in surface 
density and the velocity component along a defined slit. These variations are 
shown in Fig.~\ref{aaxlines}a and Fig.~\ref{aaxlines}b for the R801 model, and 
in Fig.~\ref{aaxlines}c and Fig.~\ref{aaxlines}d for the R991 model. 

In Fig.~\ref{aaxlines}a, it is evident that there is a density minimum at the 
central region of the R801 model. The variation in the blue curve in 
Fig.~\ref{aaxlines}b shows that the surface density on the spiral arms (local 
peaks of the blue curve) decreases as we approach the center, with a 
near-continuous decline for $\vert x\vert \lesssim 0.5 $~kpc. The $u_{\parallel}$ variation 
(orange curve) reveals that $\Delta u_{\parallel}$, measured where the slit intersects 
spiral arms also decreases closer to the center for $0.5\lessapprox \vert x\vert 
\lessapprox 1$~kpc. Near the center, $\Delta u_{\parallel}$ shows minimal variation. At 
the first crossing with a spiral arm, $\Delta u_{\parallel}\lesssim$10 km\,s$^{-1}$ (the 
extrema of the orange curve before and after the peak of the blue curve for 
$\vert x\vert \approx 1$), while for $\vert x\vert \lesssim 0.5$~kpc, $\Delta u_{\parallel}\lesssim$4 km 
s$^{-1}$.

In Fig.\ref{aaxlines}c, we see that in the R991 model, the spiral arms 
formed are even more tightly wound than in R801 (Fig.\ref{aaxlines}b).
The variation in the blue curve in Fig.\ref{aaxlines}d indicates that as we move 
along the slit toward the center, beyond the second cut with the spiral arms, 
there are small density fluctuations before the curve starts rising, eventually
peaking very close to the center. The variation in the curve highlights
the presence of faint spiral segments and a 
twin-peaks feature at $r\approx 68$~pc, morphologically similar to what is 
identified in the case of R901 with $c_s$=20~km\,s$^{-1}$. The variation in 
$\Delta u_{\parallel}$ remains small, increases along with the surface density toward 
the center, exhibiting a pattern of successive rises and falls, ultimately 
reaching a value around 10~km\,s$^{-1}$ (note the difference in scale of the 
$u_{\parallel}$ axes between Fig.~\ref{aaxlines}d and Fig.~\ref{lineR901_20_300a}b).

We conclude that in both very weak bar cases the change in the velocity parallel
to the slit ($\Delta u_{\parallel}$) is small compared to the shocks in the
main body of the bar. Specifically, shocks are not identified in the region within
the central kiloparsec of the models. The rapid fluctuations in $u_{\parallel}$ for
$\vert x\vert<1$ correspond to the region where the gas exhibits clumpiness in R991.

In summary:
\begin{itemize}

\item The responses of both R801 and R991 models suggest that, while the 
dynamics within the main bar region reflect the influence of the imposed bar 
potential, the central region is strongly influenced by the hydrodynamic 
properties of the model. In two models with similar perturbing forces as the 
R801 and R991 models, if a high surface density is initially set at the center (R991),
it persists throughout the simulation; without it (R801),
the central density remains low, lower than in its surroundings, as in the case with the 
full potential.

\item Some arising differences can also be attributed to variations in the 
relative forcing. In models with identical sound speed, the stronger 
perturbation in R901 transports more gas to the central region than the much 
weaker perturbation in R991 (Fig.~\ref{sdcentralbin}). However, qualitatively, 
models with weaker perturbing forces in the x2 region, such as R991 -- which fail 
to form nuclear or pseudo-rings -- tend to accumulate more gas near the center 
(within 100 pc) relative to the mean density in the innermost 1 kpc compared to 
models with stronger perturbations, such as R901 (cf. Fig.~\ref{aax991}b or 
Fig.~\ref{aaxlines}c with Fig.~\ref{gas901}b).

\item The presence or absence of a conspicuous central leading 
spiral depends on the forcing due to the bar component, as is indicated by its 
apparent absence in the nearly axisymmetric R801 model, while it is a 
conspicuous feature in R001 (cf. Fig.~\ref{aax801}b or Fig.~\ref{aaxlines}a with 
Fig.~\ref{fig:r1l2}b).
\end{itemize}

\section{Discussion and conclusions}
\label{sec:dis}

Gas accumulation in orbit-crowding regions arises from the non-axisymmetric
forcing in barred potentials, which drives stellar and gaseous streamlines to
converge \cite{ath92b}. In stellar models, periodic orbit families alone
predict the loci of enhanced density, but in gaseous models the sound speed,
$c_s$, critically shapes the response. At low $c_s$, gas is highly compressible,
so orbit crowding produces strong shocks and density enhancements that closely
follow stellar predictions. At higher $c_s$, pressure forces redistribute the
gas, weakening shocks, smoothing density contrasts, and shifting the dense lanes
closer to the bar’s major axis \citep{w94,wm04a,kss12}. Thus,
increasing $c_s$ reduces gas accumulation and displaces dense structures,
highlighting the growing role of pressure support in the ISM.

We carried out a series of gas response models using the
classic Ferrers bar potential. In our simulations, we did not account
for the self-gravity of the gas, nor did we include an explicit black
hole component at the center of the model.
The impact of gas self-gravity has been explicitly examined in
\citet{wh92} and subsequently in \citet{wh95}. Furthermore, \citet{wk01}
demonstrated that adopting a more realistic description of the interstellar
medium leads to resonant structures that differ qualitatively from those
obtained in simplified isothermal models. These results highlight the need for
caution when interpreting sound-speed–driven effects. Nevertheless, the present
study does not aim to resolve the detailed thermodynamical processes, but rather
to capture the general dynamical response of the gas in the central regions of
bars. In this sense, our models illustrate the principal trends and indicate the
direction in which variations in specific parameters influence the resulting
nuclear morphologies.

As \citet{Sormani_2024} indicate, the isothermal approximation
should not be interpreted literally, but rather as an effective
description that compresses a variety of complex physical processes
into a single tunable parameter. In this sense, the adopted sound
speed does not measure the true kinetic temperature of the gas;
instead, it acts as an “effective temperature” that crudely averages
over the multiphase structure of the interstellar medium, the
influence of unresolved turbulent motions, and additional forms of
pressure support, such as magnetic fields. While this treatment is
clearly idealized, its strength lies in the ability to capture, in a
controllable and transparent manner, the global dynamical trends of
gaseous flows in barred galaxies. Comparisons with more detailed
simulations indicate that the essential mechanism leading to nuclear
ring and spiral formation is robust and can be adequately reproduced within this
simplified framework.

Based on these assumptions, we draw the following conclusions:

\begin{enumerate}
 
\item In all the cases we examined, nuclear rings or nuclear spirals were 
present, with the exception of model R153, which lacked inner Lindblad 
resonances. This leads to the conclusion that the formation of nuclear 
substructures within galactic bars is consistently associated with the presence 
of an ILR (x2) region. Models without x2-flows do not have nuclear
rings, or spirals, in good agreement with what was found by \citet{ath92b}.

\item In images of galaxies, the observed nuclear rings are either round, as 
in the case of  NGC~1097, or (in most cases) rather mildly elongated, as for 
NGC~6951 or NGC~1433. In our models we encounter nuclear rings of different 
shapes. The elongation of the nuclear ring, which is approximately aligned along 
the minor axis of the bar, decreases with increasing sound speed. The most 
well-defined nuclear rings are identified in the cases with $c_s=2$~km\,s$^{-1}$,
for all models in our study. However, their elongation appears somewhat larger
than what is in general observed in real galaxies. At $c_s=10$~km\,s$^{-1}$,
only pseudo-rings are evident, while at $c_s=20$~km\,s$^{-1}$, the ring structures,
if present, as in the fiducial model R001, are smaller and rounder than in the two previous 
cases. In warm models in which they do not form, we find nuclear trailing 
spirals directly connected to the relatively straight dust lanes associated with 
the large-scale bar, in agreement with observations of galaxies such as NGC~1530.
This morphology is predominantly identified in models with an effective
sound speed of $c_s=20$~km\,s$^{-1}$.

\item We find two kinds of nuclear trailing spirals in our models: those encountered in 
models with $c_s=20$~km\,s$^{-1}$ (R010 and R192), which can be described as grand 
design, terminating in a region of low surface density, and those in 
models with $c_s=10$~km\,s$^{-1}$ (R010, R192, R901, R801, and R991), which are 
tightly wound and extend azimuthally more than $2\pi$. A particular case is the 
high $\rho_c$, warm model R901, in which a tightly wound, trailing spiral is 
formed, terminating in a featureless inner disk. Tightly wound spirals are more 
distinct in weak perturbation models (R192, R801, and R991) and less so in models 
with strong perturbing forces (R010 and R901). Comparing the responses of the same 
model but with different sound speeds, we find that the nuclear spirals become more
open, i.e., their pitch angle increases, as the sound speed increases. Typical cases are models R010 
and R192, in which we have tightly wound nuclear spirals for $c_s=10$~km\,s$^{-1}$
and open grand design patterns for $c_s=20$~km\,s$^{-1}$. Notably, 
weaker bars and nearly axisymmetric models tend to host such spirals. This is in 
agreement with the results of \citet{wm04b} and with the observations of 
\citet{mrmp03, mrmp03b}. It is also consistent with the improved 
self-consistency of tightly wound spirals when modeled under weak perturbations, 
as described by \citet{p91} and well reproduced in SPH simulations 
\citep{phgasLN}.

\item Nuclear spirals directly connected to the relatively straight dust lanes 
associated with the large-scale bar, which are observed in galaxies such as NGC~1530, are 
predominantly found in models with $c_s=20$~km\,s$^{-1}$. This does not imply
that x2-flows are absent in the corresponding stellar models. Similar morphologies
are also present in the weaker bar model R192 and the round bar model R010, with 
$c_s=10$~km\,s$^{-1}$. However, in both of these cases the nuclear spirals are 
tightly wound.

\item When a well-defined nuclear ring fails to form, tightly wound 
trailing spirals emerge within the ILR region. These spirals appear as grand design
structures in warm gas models (R001, R010, R192, and R901), while in models 
with $c_s=10$~km\,s$^{-1}$ (R010, R192, R901, R801, and R991) they are tightly 
wound. These latter spirals do not reach the system's center but terminate in a 
region of low surface density (R010 and R192) or a featureless inner disk, which 
dominates the innermost 1 kpc region (R901, R801, and R991). They are more distinct in 
weak perturbation models (R192, R801, and R991) and less so in models with strong 
perturbing forces (R010 and R901).  

\item Several studies have proposed mechanisms for the formation of leading spiral
segments in barred galaxies under specific resonance and orbital precession
conditions. In general, when two ILRs are present, the precession rate $\Omega
- \kappa/2$ decreases toward the center, generating leading nuclear spirals at
the iILR due to the positive torques exerted by the bar  \citep{c02, fc22}.
The presence of a massive central component, most effectively the presence of
a black hole, alters the shape of $\Omega - \kappa/2$, making it
increase monotonically toward the center and preventing the formation of a
leading spiral. \citet{w94} provided an analytic model for non-self-gravitating
gas in a weak bar potential, showing that damped elliptical orbits driven by
periodic forcing can produce leading or trailing spiral-like enhancements near
the first and second ILR, respectively. These results indicate that leading
spiral segments can naturally arise from the radius-dependent phase shift of
damped closed gaseous orbits under specific resonance conditions.

In our simulations, leading spiral structures are evident in models
with $c_s \lesssim 10$ km\,s$^{-1}$, particularly in those
characterized by relatively strong barred perturbations. As the
non-axisymmetric forcing weakens, the prominence of the leading
spirals correspondingly diminishes. Such spirals may coexist with
nuclear rings in the low-$c_s$ models (e.g., in the fiducial case, Fig.~\ref{fig:r1l2}), and
they can also be identified in the innermost regions of models that otherwise develop
trailing nuclear spirals (e.g., in the weak bar case, Fig.~\ref{fig:gas192}). By contrast, in
models with $c_s = 20$ km\,s$^{-1}$, the leading spiral arms are no longer discernible.
At this higher sound speed, the gas response favors the development of
more open spiral structures, with reduced winding, and the loci of
gaseous shocks are significantly displaced relative to the $x_2$
orbital domain (cf. \citealt{wm04a}). Under these conditions, the
formation of leading spirals is suppressed \citep[see also][]{kss12}.
Interestingly, in such models we consistently observe the emergence of a
“twin peaks” morphology, suggesting that a possible link between the
occurrence of leading spirals and twin peaks warrants further investigation.
 
\item In all models, while strong shocks are identified in the main body of the 
bar, the shocks in the central region are weak or negligible. Our models 
suggest that gas is funneled by the dust-lane shocks in the bar's main body into the 
ILR region. Over time, this process causes gas to accumulate either in the arcs of 
nuclear (pseudo-)rings or in the arms of tightly wound trailing spirals that 
diminish near the center. In models with high initial $\rho_c$, gas brought to 
the center compresses the existing central gas into smaller radii, further 
concentrating it near the center. This is a mechanism that promotes the
concentration of gas near the center of the system.

This process offers an explanation for the formation of twin-peak features 
identified in some models. As shocks in the main body of the bar drive more gas
into the ILR region, the compressed gas contributes to these features. This
happens either along the arcs of (pseudo-)rings that extend as continuations
of the dust-lane shocks, or close to the center in models with high central
surface densities.

\item A key conclusion is that the parameters of a gravitational potential 
alone are insufficient to predict the gas dynamics of a model. The morphology
of the gaseous response varies significantly with different sound speeds, 
highlighting the pivotal role of hydrodynamics in shaping the structure of the 
gas in the central region, in agreement with the results of \citet{w94,kss12,Sormani_2024}.

\item The models used in this study primarily highlight tendencies in the 
development of response characteristics. Nevertheless, an overall evaluation 
suggests that features in the central regions of these models align best with 
observed galaxy morphologies for an assumed sound speed of $c_s=20$~km 
s$^{-1}$. On the other hand, tightly wound spirals with an azimuthal extent  
significantly surpassing $2\pi$, such as those present in several models, are
rarely observed in galaxies. Their connection to the tightly wound nuclear
spirals described by \citet{mrmp03, mrmp03b} requires further investigation.

\end{enumerate}

\vspace{0.5cm}

\begin{acknowledgements}
We thank the anonymous referee for constructive comments on our manuscript.
PAP would like to thank MPA for the opportunity to visit and work on parts of 
this project while at the institute. This work was funded by the Sectoral 
Development Program ($\rm{O\Pi\Sigma}$ 5223471) of the Greek Ministry of Education, 
Religious Affairs and Sports, through the National Development Program (NDP) 
2021-25.  It was conducted as part of project 200/1025, 
supported by the Research Committee of the Academy of Athens.
\end{acknowledgements}

\bibliographystyle{aa}
\bibliography{bibliography}

\end{document}